\documentclass{article}
\usepackage{arxiv}
\usepackage[utf8]{inputenc}
\usepackage[T1]{fontenc}
\usepackage{microtype}
\usepackage{amsmath,amssymb,amsfonts}
\usepackage{graphicx}
\usepackage{xcolor,colortbl}
\usepackage{threeparttable,tabularx,booktabs,array,multirow}
\usepackage[section]{placeins} 
\usepackage{steinmetz}
\usepackage[numbers,sort&compress]{natbib}
\usepackage{url}
\usepackage{hyperref}
\graphicspath{{./images/}}

\renewcommand{\shorttitle}{Mixed-effects random forests for multi-cohort predictions}
\renewcommand{\headeright}{Preprint}
\renewcommand{\undertitle}{Preprint}

\title{Advancing mixed-effects random forests for predictions based on multi-cohort data}
\author{%
  \textbf{Jiumeng Zhang}$^{1}$ \qquad \textbf{Jordan Behrendt}$^{2}$ \qquad \textbf{Tanja Schultz}$^{2}$\\[0.25em]
  \textbf{Tanja Vrijkotte}$^{3}$ \qquad \textbf{Siyu Zhou}$^{3}$\\[0.25em]
  \textbf{Iris Pigeot}$^{1}$ \qquad \textbf{Claudia B{\"o}rnhorst}$^{1}$\\[1em]
  \small $^{1}$Leibniz Institute for Prevention Research and Epidemiology -- BIPS, Bremen, Germany\\
  \small $^{2}$University of Bremen, Bremen, Germany\\
  \small $^{3}$Amsterdam UMC, Amsterdam, Netherlands\\[0.6em]
  \small Correspondence: Claudia B{\"o}rnhorst (\href{mailto:boern@leibniz-bips.de}{boern@leibniz-bips.de})
}
\date{}
\hypersetup{
  hidelinks,
  pdftitle={Advancing mixed-effects random forests for predictions based on multi-cohort data},
  pdfauthor={Jiumeng Zhang, Jordan Behrendt, Tanja Schultz, Tanja Vrijkotte, Siyu Zhou, Iris Pigeot, Claudia Bornhorst},
  pdfkeywords={Covariance structure, Mixed-effect models, Mixed-effects machine learning, Predictive modeling, Random effects}
}

\begin{document}
\maketitle

\begin{abstract}
Pooling cohorts could extend the age coverage and may improve predictions of long-term health outcomes by sharing information across time. Mixed-effects machine learning (ME-ML) combines mixed-effects models with machine learning methods to account for dependencies while flexibly capturing complex patterns in correlated data. However, these ME-ML methods are limited in their applicability in complex settings such as multi-cohort data. To improve the predictions of individual-level health trajectories beyond the observed period, we propose mixed-effects random forests for multi-cohort data (MERF$_{multi-cohort}$) which extends previous mixed-effects random forest (MERF) in three respects: (i) better separation of population-level and subject-specific variation; (ii) flexibility in correcting systematic discrepancies in fixed-effects estimates to improve covariance parameter estimation; and (iii) flexible specification of variance–covariance structures for random effects and residuals. We evaluate the proposed method against linear regression, linear mixed models, random forests, and standard MERF through simulation studies and in four real-data scenarios predicting BMI z-scores in children and adolescents using single- and multi-cohort data. Simulation results show that MERF$_{multi-cohort}$ improves covariance parameter estimation under various variance–covariance structures while maintaining predictive performance comparable to MERF. Improvement in predictive accuracy is most obvious when residual variances differ across study waves, indicating the benefit of modeling measurement heterogeneity. In real-data applications, MERF$_{multi-cohort}$ also shows improved predictive performance over MERF by flexibly capturing complex patterns and accommodating cohort heterogeneity. Additionally, adding data from similar cohorts covering extended age ranges enhances the accuracy of future individual-level predictions if the heterogeneity in multi-cohort data is adequately accounted for.

\end{abstract}


\section{Introduction}\label{sec1:intro}

Early prediction of chronic diseases is crucial for informing personalized interventions that promote health and well-being across the life course. Although observing developmental processes over time is ideal for understanding mechanisms of diseases with long latency periods, in practice many cohort studies cover only a limited age range, making it difficult to predict (markers of) diseases long before an outbreak.

Pooling multiple cohorts spanning different but overlapping stages in life could extend the observation periods \citep{b1,b3}. We anticipate that sharing information across cohorts may be an efficient way to improve the predictions of individual-level health trajectories. However, joint analysis of multi-cohort data inevitably adds complexity because of the heterogeneity across studies and the hierarchical data structure (repeated measurements nested within subjects, subjects nested within cohorts). So far, there is a striking lack of analytical techniques that can adequately predict health outcomes over extended age spans based on multi-cohort data.

In predictive modeling, machine learning (ML) methods such as random forests (RF) \citep{b4} are widely applied because of their flexibility in capturing non-linear relationships and interactions in complex data. However, they assume that observations are independent and identically distributed (i.i.d). In single-cohort settings, this assumption is already violated because repeated observations are nested within subjects, and subjects may be further nested within larger units (e.g., families or clinics). (Generalized) Linear Mixed Models (LMMs) are widely used in epidemiology to accommodate longitudinal, clustered data but rely on parametric assumptions and require predefined interactions among predictors. Mixed-effects machine learning (ME-ML) methods, which combine (Generalized) LMMs with ML, have shown improved prediction performance in single-cohort settings by leveraging the strengths of both approaches while mitigating limitations \citep{Ngufor}. However, these ME-ML methods remain underexplored in multi-cohort settings with more complex hierarchical structures. 

To fill this gap, we propose multi-cohort mixed-effects random forests (MERF$_{multi-cohort}$), a new approach that combines LMMs and RF to improve the predictions of individual-level health trajectories beyond the actual individual measurement period based on multi-cohort data. The novelty of our method lies in extending MERF to accommodate complex covariance structures to allow more flexibility in modeling multi-cohort data. 

The structure of this paper is as follows. In Section 2, we present a motivating example. In Section 3, we introduce the general ME-ML framework and our MERF$_{multi-cohort}$ approach. In Section 4, we present simulation studies that assess the performance of the proposed method. In Section 5, we describes the experimental setup for illustrating our approach using a real-data example of predicting BMI z-scores in children based on either single-cohort or multi-cohort data. Section 6 presents the empirical findings and interpretation of the results. Section 7 provides a discussion and conclusions.

 \section{Motivating Example}\label{sec2:Motivating Example}

For illustration, we consider the task of predicting health outcomes (e.g., BMI z-scores) at future ages in children from information collected in cohort studies, although the proposed approach is more generally applicable to longitudinal health outcomes across later life stages. Suppose we have one target cohort and aim to predict health outcomes for its subjects at ages beyond the period covered in that cohort. We anticipate that augmenting the training data of the target cohort with data from other cohorts covering extended age spans could improve the prediction of individual-level health outcome trajectories in the target cohort. Therefore, we aim to investigate whether the accuracy of individual-level health outcome predictions can be improved by complementing data from the target cohort with data from other cohorts, particularly in cases where the target cohort is limited in the age range covered. 

One challenge when working with multi-cohort data is measurement heterogeneity, which arises when different measurement tools are used in different studies or even across waves within the same study \citep{b3}. Although the variables are often harmonized using simple algorithmic transformations (e.g., combining categories, recoding), this does not fully resolve the measurement differences, which still need to be considered in modeling. Additionally, multi-cohort data have at least three hierarchical levels (i.e., repeated measurements nested within subjects, and subjects nested within cohorts). The hierarchical structure can be even more complex when additional units, such as clinics or families, exist between the subject and cohort levels. Such multi-level hierarchical structure introduces multiple sources of variation that need to be accounted for in the analysis. Therefore, advanced analytical techniques are required to accommodate the complexities of multi-cohort data.

\section{Methods}\label{sec3:Methods}

\subsection{Mixed-effects Models}

(Generalized) Linear Mixed Models (LMMs) were developed for data with correlated observations. For continuous outcomes, they can be formulated as \citep{Harville1977}:
\begin{equation}
  \mathbf{y}_i = \mathbf{X}_i\,\boldsymbol{\beta} + \mathbf{Z}_i\,\boldsymbol{\nu}_i + \boldsymbol{\varepsilon}_i,\qquad i \in \{1,\ldots,n\},
\label{eq:lmm}
\end{equation}
where $\mathbf{y}_i$ is the $(n_i\times 1)$ vector (vector of BMI z-scores of subject $i$ in our motivating example), $n_i$ is the number of measurements taken on subject $i$, $\mathbf{X}_i$ is the $(n_i\times p)$ matrix of $p$ predictors collected on subject $i$, $\boldsymbol{\beta}$ is the $(p\times 1)$ vector of fixed effects parameters, $\mathbf{Z}_i$ is a $(n_i\times q)$ design matrix, $\boldsymbol{\nu}_i$ is a $(q\times 1)$ vector of unobserved random effects for subject $i$ with $\boldsymbol{\nu}_i\sim \mathcal{N}(\mathbf{0},\,\boldsymbol{D}_i)$, and $\boldsymbol{\varepsilon}_i$ is the $(n_i\times 1)$ vector of normally distributed random errors with $\boldsymbol{\varepsilon}_i\sim \mathcal{N}(\mathbf{0},\,\boldsymbol{R}_i)$.

By choosing suitable covariance structures $\boldsymbol{D}_i$ and $\boldsymbol{R}_i$ for the residuals and random effects, LMMs can account for dependencies in correlated data.

\subsection{Mixed-effects Machine Learning (ME-ML)}
ME-ML approaches are built on the idea of integrating the random effects structure of LMMs into flexible ML methods and are typically formulated as \citep{Ngufor,Kilian}:
\begin{equation}
  \mathbf{y}_i = f(\mathbf{x}_i) + \mathbf{Z}_i\,\boldsymbol{\nu}_i + \boldsymbol{\varepsilon}_i, \qquad i \in \{1,\ldots,n\}.
\label{eq:MEml}
\end{equation}
The fixed effects part $\mathbf{X}_i\,\boldsymbol{\beta}$ in LMMs is replaced by a more general unknown function $f$ to capture complex patterns. In theory, the unknown function $f$ could be modeled by any ML approach. ME-ML approaches typcially alternate between estimating the fixed effects with a ML model, and estimating the random effects $\boldsymbol{\nu}_i$ with a conventional LMM until a ceratin prespecified convergence criterion is reached.

\subsection{Mixed-effects Random Forests (MERF)}\label{s:MERF}
Among the broad class of ME-ML methods, we focus particularly on MERF because RFs are applicable to a wide range of prediction problems and are generally recognized for their accuracy with few parameters to tune \citep{b6}. Comprehensive reviews of existing MERF approaches are provided by \citet{MERFreview} and Capitaine et al.~\citep{Capitaine2021}. In general, most MERF approaches follow the formulation \citep{Hajjem2014}:
\begin{equation}
\mathbf{y}_i = f_{RF}(\mathbf{x}_i) + \mathbf{Z}_i\,\boldsymbol{\nu}_i +  \boldsymbol{\varepsilon}_i,
\end{equation}
where $f_{RF}$ denotes the RF for the fixed effects, $\mathbf{Z}_i$ is a $(n_i\times q)$ design matrix, $\boldsymbol{\nu}_i$ is a $(q\times 1)$ vector of unobserved random effects for subject $i$ with $\boldsymbol{\nu}_i\sim \mathcal{N}(\mathbf{0},\,\boldsymbol{D})$, and $\boldsymbol{\varepsilon}_i$ is the $(n_i\times 1)$ vector of normally distributed random errors with $\boldsymbol{\varepsilon}_i\sim \mathcal{N}(\mathbf{0},\,\sigma_\varepsilon^2\,\mathbf{I}_{n_i})$.
MERF estimates the fixed effects using a random forest $f_{RF}(\mathbf{x}_i)$ and alternates between
estimating the fixed effects and predicting the random
effects with a conventional LMM. In the first iteration, random effects are initialized to be 0. At iteration $t$ $(t>1)$, the RF is fitted to the outcome after removing the current estimated random contribution
\begin{equation}
\mathbf y_i^{*(t)}
=
\mathbf y_i
-
\mathbf Z_i
\hat{\mathbf \nu}_i^{(t-1)},
\end{equation}
giving
\begin{equation}
\hat f_{RF}^{(t)}
=
RF
\left(
\mathbf X,
\mathbf Y^{*(t)}
\right).
\label{eq:merf_rfstep}
\end{equation}
The predictions from $\hat f_{RF}^{(t)}$ are subsequently treated as the fixed effects contribution $\mathbf{\hat{y}}_{\text{fixed}}^{(t)}$ in the LMM:
\begin{equation}
\mathbf{y}_{i} = \mathbf{\hat{y}}_{\text{fixed}}^{(t)}  + \mathbf{Z}_{i}\boldsymbol{\nu}_{i} + \boldsymbol{\varepsilon}_{i}.
\end{equation}

This formulation has two limitations which make it too restrictive in multi-cohort settings. 

First, RF could capture the pattern of the true fixed-effects function well, but systematic discrepancies in location and/or scale may still remain in the fixed effect predictions. Let
$f_0(\mathbf X_{ikj})$ denote the true fixed effect 
function for measurement $j$ of subject $i$ in cohort $k$. The true data-generating model could be written as
\begin{equation}
y_{ikj}
=
f_0(\mathbf X_{ikj})
+
\mathbf Z_{ikj}\mathbf \nu_{ik}
+
\epsilon_{ikj}.
\label{eq:true_model}
\end{equation}
At each MERF iteration, the RF predictions enter the LMM directly as the fixed effect component, so that the LMM is fitted to the residuals after subtracting
$\mathbf{\hat{y}}_{\text{fixed}}^{(t)}$ from the outcome. The
fixed effect function estimation error is
\begin{equation}
d_{ikj}
=
f_0(\mathbf X_{ikj})
-
\hat f_{\mathrm{RF}}(\mathbf X_{ikj}).
\label{eq:fixed_function_error}
\end{equation}
Then the RF residual is
\begin{align}
y_{ikj}
-
\hat f_{\mathrm{RF}}(\mathbf X_{ikj})
&=
f_0(\mathbf X_{ikj})
-
\hat f_{\mathrm{RF}}(\mathbf X_{ikj})
+
\mathbf Z_{ikj}\mathbf \nu_{ik}
+
\epsilon_{ikj}
\\
&=
d_{ikj}
+
\mathbf Z_{ikj}\mathbf \nu_{ik}
+
\epsilon_{ikj}.
\label{eq:residual_decomposition}
\end{align}
Hence, unless $d_{ikj}$ is negligible, the RF residuals which are subsequently used to estimate the random effect and residual covariance structures contain not only the intended components $\mathbf Z_{ikj}\mathbf \nu_{ik}+\epsilon_{ikj}$, but also systematic error in the estimated fixed-effect function $\hat f_{RF}$. 

When data from multiple cohorts are pooled, this issue can be particularly relevant for estimating parameters of  the random effects and residuals variance-covariance matrix (i.e., $\boldsymbol{D}_i$ and $\boldsymbol{R}_i$). For example, suppose cohort 1 covers ages 5-15 years and cohort 2 ages 6-25 years. The pooled sample then contains more information in the overlapping age range 6-15 years than at the upper end of the age distribution (>15 years), where only cohort 2 contributes. If the true regression function varies nonlinearly with age, the fixed-function estimation error $d_{ikj}$ may therefore differ across the age range. In such a setting we may have, for example, $d_{i1j}\approx 0$ for measurements in cohort 1 with more age support, but $d_{i2j}$ deviates from 0 systematically for later-age measurements in cohort 2. The fixed effect function estimation error then has both cohort and longitudinal structure which may contaminate the estimation of the  $\boldsymbol{D}_i$ and $\boldsymbol{R}_i$.

Second, previous MERF approaches are limited in the flexibility of their covariance structures for $\boldsymbol{D}_i$ and $\boldsymbol{R}_i$ since they were mainly applied to settings with rather simple correlation structures (e.g., clustered data, single cohorts). Most existing approaches assume homogeneity of the within-subject residual variance and assume that no serial correlation exists between the within-subject residuals (i.e., $\mathbf{R}_i$ $=$ $\sigma_\varepsilon^2\,\mathbf{I}_{n_i}$). As mentioned in our motivating example, when working with multi-cohort data, the measurement heterogeneity should be considered in modeling. Additionally, empirical studies \citep{Grajeda2016,Liu2025,Vemuri2014} suggest that serial correlation sometimes remains even after accounting for random intercepts/slopes. In such cases, the assumption of no correlation between the within-subject residuals is too restrictive and a more flexible correlation structure for the residuals is required. 

To adapt MERF to settings requiring more complex variance-covariance structures, Capitaine et al.~\citep{Capitaine2021} proposed stochastic mixed effects random forests (SMERF) and stochastic random effects expectation maximization forests (SREEMforest). Both SMERF and SREEMforest extend MERF by incorporating a stochastic serial correlation effects which allows the covariance structure to vary over time: 
\begin{equation}
\mathbf{y}_i = f_{RF}(\mathbf{x}_i) + \mathbf{Z}_i\,\boldsymbol{\nu}_i + \boldsymbol{\omega}_i + \boldsymbol{\varepsilon}_i,
\label{eq:SMERF}
\end{equation} where $\boldsymbol{\omega}_i$ is a stochastic process (e.g.,  Ornstein-Uhlenbeck process) used to model serial correlation for individual $i$. 

While useful in many longitudinal settings, the MERF approach suggested by Capitaine et al.~\citep{Capitaine2021} still  presents methodological limitations that restrict its applicability to multi-cohort data. The stochastic process requires a sufficient number of measurements per subject to obtain stable estimates. In many cohort studies, however, each subject has only a limited number of measurements despite large sample sizes. Additionally, the estimated parameters in the variance-covariance matrix for random effects are assumed to be constant across all subjects (i.e., $\mathbf{D}_i$ $=$ $\mathbf{D}$) in the MERF approach by Capitaine et al.~\citep{Capitaine2021}. In multi-cohort settings, some cohorts might have greater between-subject variability when the sample is from a more heterogeneous population. In such cases, ignoring the heterogeneity in the variance-covariance matrix will either overestimate between-subject variability in the less heterogeneous cohorts or underestimate it in the more heterogeneous cohorts, which can lead to poor prediction of individual-level outcome trajectories \citep{Daniels2003}.

\subsection{Mixed-effects Random Forest for Multi-cohort Data (\texorpdfstring{MERF$_{multi-cohort}$}{MERF multi-cohort})}
Our MERF$_{multi-cohort}$  generalizes MERF in three respects: (i) we use subject-level rather than observation-level out-of-bag (OOB) prediction to reduce the extent to which subject-specific variation is absorbed by the RF; (ii) we relax the assumption that the RF estimate enters the mixed-effects model directly with unit coefficient by introducing an affine calibration of the fixed effect contribution; (iii) we allow more flexible variance--covariance structures for both random effects and residuals estimated as parts of the LMM.

\subsubsection{Subject-level cross-fitting of the RF component}
\label{section:subjectOOB}
In standard RF with repeated measurements, an observation can be OOB for a particular tree while other observations from the same subject are still in-bag, which can make the observation-level OOB prediction error too optimistic for longitudinal data \citep{MERFreview}. 

To reduce the subject-specific variation absorbed by the RF, we construct RF predictions using $H$-fold cross-fitting at the subject level. $\mathcal I_1,\ldots,\mathcal I_H$ are mutually exclusive sets of subjects in the training data. For a subject $i\in\mathcal I_h$, the RF prediction is obtained from a forest $
\hat f^{(-\mathcal I_h)}
(
\mathbf X_{i}
)$ trained using subjects outside $\mathcal I_h$.

In this way, no observation from subject $i$ contributes to the RF used to construct that subject's fixed effect prediction, which provides a better separation between population-level variations used to estimate the fixed effect component and the subject-specific variations subsequently estimated through the random effects.

\subsubsection{Calibrated fixed effect component}
At iteration $t$, let
\[
\hat{\mathbf f}_i^{(t)}
=
\left[
\hat f^{(t)}(\mathbf X_{i1}),
\ldots,
\hat f^{(t)}(\mathbf X_{in_i})
\right]^\top
\]
denote the RF predictions for subject $i$. We propose the MERF approach with calibrated fixed effect component
\begin{equation}
\mathbf y_i
=
\mathbf W_i^{(t)}
\boldsymbol\beta
+
\mathbf Z_i\mathbf \nu_i
+
\boldsymbol\epsilon_i,
\label{eq:RFconditionlmm}
\end{equation}

where
\[
\mathbf W_i^{(t)}
=
\begin{bmatrix}
\mathbf 1_i &
\hat{\mathbf f}_i^{(t)}
\end{bmatrix},
\qquad
\boldsymbol\beta
=
\begin{pmatrix}
\alpha\\
\gamma
\end{pmatrix}.
\]

Conditional on a given RF prediction, the original MERF is nested within the proposed model as the restricted case when $\alpha=0$ and $\gamma=1$.

The calibration does not introduce additional nonlinear structure, it only provides more flexibility in correcting the systematic location and scale discrepancy in the RF estimate.

\subsubsection{Flexible random effect and residual covariance structures}
To account for between-cohort variability, cohort-level random effects can be specified in the LMM to directly model the hierarchical structure (measurement $j$ nested within subject $i$ and subject $i$ nested within cohort $k$):
\begin{equation}
y_{ij} = \alpha + \mathbf{Z}_{1,ij}\boldsymbol{\nu}_{i} + \mathbf{Z}_{2,ij}\boldsymbol{u}_{k(i)} + \varepsilon_{ij},
\label{eq:nestedD}
\end{equation}
\[\boldsymbol{\nu}_i \sim \mathcal{N}(\mathbf{0},\,\boldsymbol{D}_{\nu})\]

where $k(i)$ denotes the cohort k containing subject $i$. $\mathbf{u}_{k(i)}$ is the cohort-level random effects vector, $\mathbf{\nu}_{i}$ is the subject-level random effects vector,
$\varepsilon_{ij}$ is the within-subject error. 

To account for variation across cohorts in the magnitude and structure of between-subject heterogeneity, for $K$ cohorts with possibly different random effect structures per cohort (e.g., random intercept in one cohort, random intercept and random slope of age in other cohorts), we use an indicator $c_{k}(i)$ to denote subject $i$ in cohort $k$. Let $\mathbf{z}_{ij}$ denote the design vector
associated with the random effects for measurement $j$ of subject $i$ and $\boldsymbol{\nu}_{ik}$ is the vector of random effects for subject $i$ in cohort $k$. The subject-level random effects component for subject $i$ in cohort $k$ is: 
\[
\sum_{k=1}^{K} c_{k}(i)\,\mathbf{z}_{ij}^{T}\boldsymbol{\nu}_{ik},
\label{eq:Dcohort}
\]
where 
\[
\mathbf \nu_{ik}
\sim
N(\mathbf 0,\mathbf D_{k(i)}).\]

This allows each cohort to have its own set of parameters in the variance-covariance matrix $\mathbf D_{k(i)}$ for subject-level random effects. This way, the estimated variance parameters are not constrained to be equal, and different random effects structures can also be specified for each cohort. The random effects for different cohorts are uncorrelated and therefore the variance-covariance matrix of subject-level random effects is block-diagonal across cohorts:
\[
\mathbf{D} =
\begin{pmatrix}
\mathbf{D}_{1} & \mathbf{0} & \cdots & \mathbf{0} \\[6pt]
\mathbf{0} & \mathbf{D}_{2} & \cdots & \mathbf{0} \\[6pt]
\vdots & \vdots & \ddots & \vdots \\[6pt]
\mathbf{0} & \mathbf{0} & \cdots & \mathbf{D}_{K}
\end{pmatrix}.
\]
The choices are not restricted to the specification presented here. In principle, more complex hierarchical or cross-classified data structures could also be accommodated through the specification of the random effects structure depending on the specific data settings.

To account for measurement heterogeneity across different study waves and the within-subject temporal correlation not captured by random effects, we assume the residual variance-covariance matrix to be

\begin{equation}
\mathbf{R}_i
=
\mathbf{S}_i
\mathbf{C}_i(\rho)
\mathbf{S}_i,
\end{equation}

where

\[
\mathbf{S}_i
=
\operatorname{diag}
\left(
\sigma_{g_{i1}},
\sigma_{g_{i2}},
\ldots,
\sigma_{g_{in_i}}
\right)
\]

Here $g_{ij}$ ($j\in \{1,\ldots,n_i\}$) denotes the study wave associated with observation $j$ of subject $i$, and \(\sigma_g^2\) denotes the residual variance for wave \(g\). Therefore $\mathbf{S}_i$ contains the study wave-specific residual standard deviations, and
\(\mathbf{C}_i(\rho)\) is a correlation
matrix with continuous autoregressive model of order 1 (CAR(1)) structure:

\[
\left[\mathbf{C}_i(\rho)\right]_{jj'}
=
\rho^{\left|t_{ij}-t_{ij'}\right|},
\]
where $\rho$ is the correlation parameter which indicates how quickly correlation to decrease as time interval between measurements $\left|t_{ij}-t_{ij'}\right|$ increases. This allows the correlation between two observations decreases in magnitude as the time interval increases and accommodates the irregular measurement intervals typical for multi-cohort data (since measurement schedules across cohorts are not aligned in general).

The greater flexibility in the specification of the random effects structure and residual variance--covariance structure helps to better capture between- and within-subject variation in multi-cohort data.

\subsection{Implementation of \texorpdfstring{MERF$_{multi-cohort}$}{MERF multi-cohort}}

The proposed MERF$_{multi-cohort}$ algorithm is shown in Table~\ref{t:MERF_cohort}. The random effects $\mathbf{\hat{\nu}}_{i}$ are initialized to be 0 at the beginning since both fixed effects $\mathbf{y}^{\text{fixed}}$ and random effects $\mathbf{\nu}_{i}$ are unknown. The adjusted outcome $\mathbf y^{*(t)}$is constructed by subtracting the current random effects prediction from the original outcome, which is intended to contain mainly the population-level variation for the RF to capture. As detailed in ~\ref{section:subjectOOB}, we fit RF using subject-level folds to the adjusted outcome $\mathbf y^{*(t)}$ and estimate the fixed effect component using subject-level OOB RF predictions. 

Given RF predictions, calibration parameters $\boldsymbol\beta$ ($\alpha$ and $\gamma$), and the covariance parameters $\boldsymbol\psi$ contained in
$\mathbf D_k$ and $\mathbf R_{i}$ are jointly estimated within the LMM. Maximum likelihood rather than restricted maximum likelihood is used during the iterative procedure because the fixed effect components $\hat{\mathbf f}^{(t)}$ changes across iterations. The random effects predictions are subsequently updated based on the LMM parameter estimates. 
This provides greater flexibility in the specification of the random effects structure and residual variance--covariance structure. In principle, a wide range of variance--covariance structures typical of LMMs can be considered in MERF$_{multi-cohort}$ depending on the data characteristics and the underlying scientific question. Additionally, the likelihood evaluated during the LMM fit is used to monitor iteration to iteration stabilization.
At each iteration, $\tilde{\ell}_t$ denotes the log-likelihood of the fitted LMM evaluated at the estimated coefficients. The relative change in the log-likelihood is then calculated to assess convergence. The alternation between estimating fixed effects by RF and predicting random effects by LMM continues until the relative change in the log-likelihood is less than a prespecified tolerance value or the prespecified maximum number of iterations is reached. In similar EM-type alternating procedures for ME-ML \citep{Sela2012, Krennmair2022}, tolerance values between $\mathbf{10}^{\text{-3}}$ and $\mathbf{10}^{\text{-5}}$ are used to track the change of absolute/relative change in the log-likelihood to balance stability and efficiency. The specific tolerance value used as default in our implementation is $\mathbf{10}^{\text{-4}}$ with maximum 25 iterations. The R code implementing the proposed method is available in the online Supplementary Material.

\begin{table}[h]
\centering
\caption{MERF$_{\text{multi-cohort}}$ algorithm}
\label{t:MERF_cohort}
\begin{tabular}{|p{0.10\linewidth}|p{0.82\linewidth}|}
\hline

\multicolumn{2}{|l|}
{\textbf{Algorithm: Multi-cohort Mixed-effects Random Forest}} 
\\ \hline

\textbf{Input}
&
Longitudinal data
$\{(\mathbf x_{ij},y_{ij}),\,
i=1,\ldots,N,\;
j=1,\ldots,n_i\}$
\\ \hline

\textbf{Output}
&
Estimated random forest $\hat f_{RF}$,
calibration parameters
$\hat{\alpha}$ and $\hat{\gamma}$,
covariance parameter estimates
$\hat{\boldsymbol\psi}$,
and random-effect predictions
$\hat{\mathbf \nu}_i$
\\ \hline

$t=0$
&
Initialize
\[
\hat{\mathbf \nu}_i^{(0)}
=
\mathbf 0,
\qquad
i=1,\ldots,N.
\]
\\ \hline
$t=$ \\ $t+1$
&
\begin{minipage}[t]{0.95\linewidth}

Iterate through the following steps until converge:

\medskip

\textbf{(a) Construct the adjusted outcome}

\[
\mathbf y_i^{*(t)}
=
\mathbf y_i
-
\mathbf Z_i
\hat{\mathbf \nu}_i^{(t-1)}.
\]

\textbf{(b) Estimate the fixed effect component $\hat{\mathbf f}^{(t)}$ using subject-level cross-fitting}

For each subject-level fold $h=1,\ldots,H$, fit RF
$\hat f^{(-h,t)}_{\mathrm{RF}}$
to the adjusted outcomes using subjects outside fold $h$.
For observations belonging to fold $h$, obtain

\[
\hat{\mathbf f}_i^{(t)}
=
\hat f^{(-h,t)}_{\mathrm{RF}}(\mathbf X_i).
\]

\textbf{(c) Fit the LMM}
\[
\mathbf y_i
=
\mathbf W_i^{(t)}
\boldsymbol\beta
+
\mathbf Z_i\mathbf \nu_i
+
\boldsymbol\epsilon_i,
\]

where

\[
\mathbf W_i^{(t)}
=
\begin{bmatrix}
\mathbf 1_i &
\hat{\mathbf f}_i^{(t)}
\end{bmatrix},
\qquad
\boldsymbol\beta
=
\begin{pmatrix}
\alpha\\
\gamma
\end{pmatrix}.
\]
with
\[
\mathbf \nu_i
\sim
\mathcal N
\left(
\mathbf 0,
\mathbf D_{k}
\right),
\qquad
\boldsymbol\epsilon_i
\sim
\mathcal N
\left(
\mathbf 0,
\mathbf R_i
\right),
\]

estimate $\alpha$, $\gamma$, and the covariance parameters
$\boldsymbol\psi$.

\textbf{(d) Update the random effect predictions}

\[
\hat{\mathbf \nu}_i^{(t)}
=
\hat{\mathbf D}_{k}
\mathbf Z_i^\top
\hat{\mathbf V}_i^{-1}
\left[
\mathbf Y_i
-
\hat{\alpha}_t\mathbf 1_i
-
\hat{\gamma}_t
\hat{\mathbf f}_i^{(t)}
\right],
\]

where

\[
\hat{\mathbf V}_i
=
\mathbf Z_i
\hat{\mathbf D}_{k}
\mathbf Z_i^\top
+
\hat{\mathbf R}_i.
\]

\textbf{(e) Assess convergence}

Compute the relative change in the log-likelihood

\[
\Delta_\ell^{(t)}
=
\frac{
\left|
\tilde{\ell}_t-\tilde{\ell}_{t-1}
\right|
}{
1+
\left|
\tilde{\ell}_{t-1}
\right|
}.
\]

Stop when

\[
\Delta_\ell^{(t)}
<
\varepsilon,
\]

or when the prespecified maximum number of iterations is reached.\\

\end{minipage}
\\ \hline

\end{tabular}
\end{table}


\subsubsection{Prediction}

After convergence, fixed effect predictions for new subjects (i.e., subjects with no previous observations seen during model training) are obtained from the final RF and calibration parameters,
\begin{equation}
\hat y_{\mathrm{fixed}}(\mathbf x_{ij})
=
\hat\alpha
+
\hat\gamma
\hat f_{\mathrm{RF}}(\mathbf x_{ij}).
\end{equation}

When only subject-level random effects are included, for subject $i$ with previous observations included in the training set, the prediction consists of the fixed effect predictions and the BLUP of the subject-specific random effects:
\begin{equation}
{
\hat y_{ij}^{\mathrm{seen}}(\mathbf x_{ij})
=
\hat\alpha
+
\hat\gamma
\hat f_{\mathrm{RF}}(\mathbf x_{ij})
+
\mathbf Z_{ij}
\hat{\mathbf \nu}_i.
}
\end{equation}

When cohort-level random effects are included in addition to subject-level random effects, for subject $i$ with previous observations included in the training set, the prediction additionally incorporates the BLUP of the cohort-specific random effects:

\begin{equation}
{
\hat y_{ij}^{\mathrm{seen}}(\mathbf x_{ij})
=
\hat\alpha
+
\hat\gamma
\hat f_{\mathrm{RF}}(\mathbf x_{ij})
+
\mathbf Z_{1,ij}
\hat{\mathbf \nu}_i+ \mathbf Z_{2,ij}
\hat{\mathbf u}_{k(i)},
}
\end{equation}

where $k(i)$ denotes the cohort k containing subject $i$.

\section{Simulation Study}
\subsection{Simulation Design}
 In Section~\ref{s:setup}, we apply the proposed method to predict BMI z-score trajectories because BMI is easy to measure and widely available in many cohort studies. However, the proposed MERF$_{multi-cohort}$ could also be applied to other health outcomes that are more difficult or costly to measure, where leveraging previous measurements to obtain future predictions may be particularly beneficial. For example, measurement of fasting insulin requires blood collection under fasting status which is challenging in children.

We simulated longitudinal data from two child cohorts, with repeated measurements of fasting insulin concentrations on the natural-log scale as outcome. To mimic the design of real multi-cohort data, the simulated two-cohort data have:
\begin{itemize}
    \item different design features (e.g., number of subjects, age coverage at baseline, measurement-wave schedules);
    \item irregular actual measurement times around planned measurement-waves;
    \item different numbers of repeated measurements within cohorts due to dropout;
    \item differences in covariate distribution.
\end{itemize}

For subject $i$ in cohort $k$ at measurement $j$, the outcome (fasting insulin concentration on the natural-log scale) was generated as

\begin{equation}
y_{kij}
=
f(\mathbf{x}_{kij}, t_{kij})
+ \mathbf{z}_{1,kij}\boldsymbol{u}_{k}
 + \mathbf{z}_{2,kij}\boldsymbol{\nu}_{i} 
 + \varepsilon_{kij}, 
\end{equation}

where $t_{kij}$ denotes the age of subject $i$ from cohort $k$ at measurement $j$, $f(\cdot)$ is a nonlinear fixed effects function of age and  covariates $\mathbf{x}_{kij}$ (details provided in Appendix ~\ref{simulation}), $\boldsymbol{u}_{k}$ denotes cohort-level random effects, $\boldsymbol{\nu}_{i}$ denotes subject-level random effects, and $\varepsilon_{kij}$ denotes the within-subject residual error.

\begin{figure}[t]
  \centering

  \includegraphics[width=\linewidth]{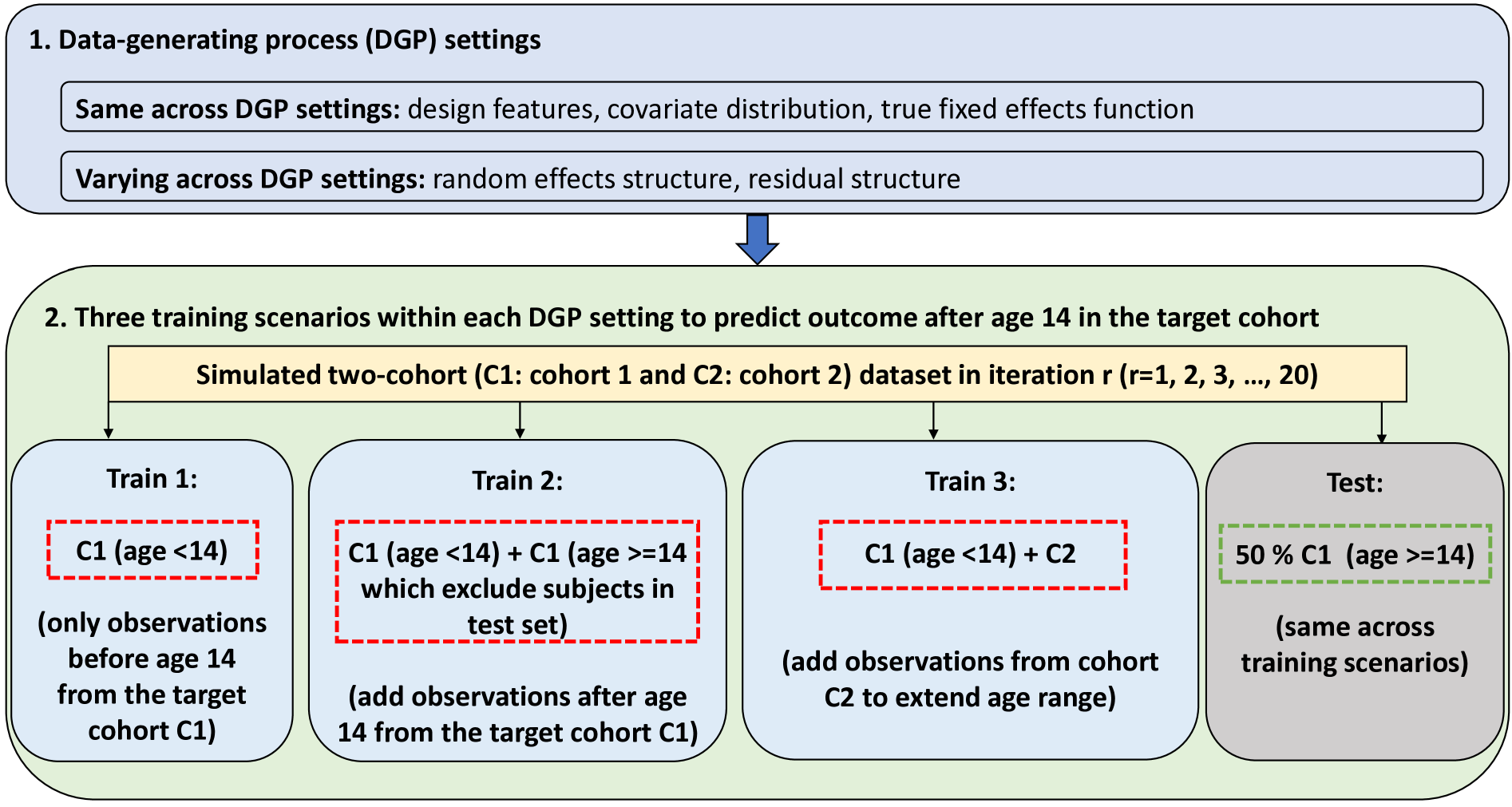}

  \includegraphics[width=\linewidth]{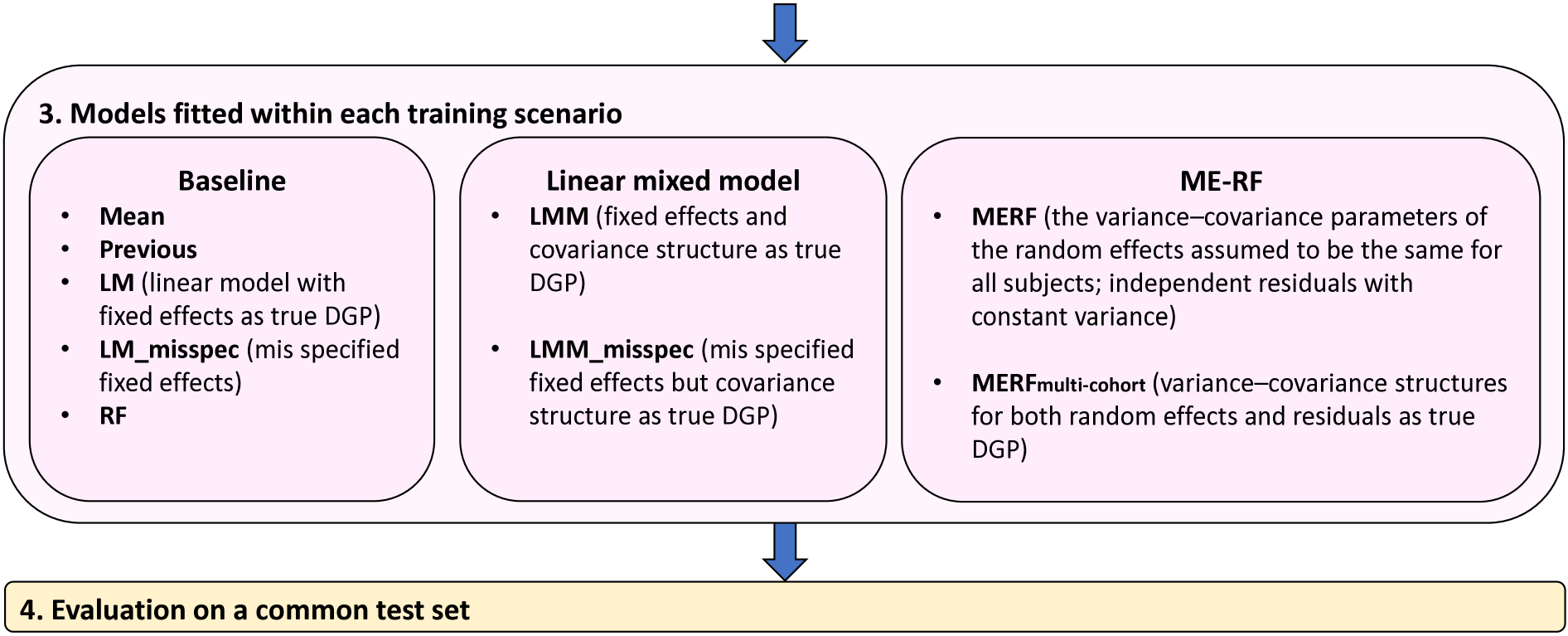}

  \caption{Simulation design}
  \label{f:simulation_design}

  \vspace{0.5em}

  \begin{minipage}{\linewidth}
    \footnotesize
    \noindent\textit{Note:}
    Cohort C1 consists of 8,000 subjects with baseline
    ages uniformly distributed between 4.5 and 7 years,
    while Cohort C2 consists of 15,000 subjects with
    baseline ages uniformly distributed between 2 and
    10 years. Four measurement waves are scheduled at
    0, 3, 6, and 11 years after baseline for C1 and
    at 0, 2, 6, and 13 years after baseline for C2.
    Actual measurement ages are obtained by adding
    independent uniform deviations between $-0.5$
    and $0.5$ years to the scheduled ages and are
    rounded to one decimal place. In both cohorts subjects have a dropout probability of 0.30 at each follow-up wave.
  \end{minipage}

\end{figure}

\begin{table}[t]
\centering
\begin{threeparttable}
\caption{Random effects structures considered in different DGP settings}
\label{tab:random_effect_structures}

\normalsize
\renewcommand{\arraystretch}{1.35}
\setlength{\tabcolsep}{5pt}

\begin{tabularx}{\textwidth}{
>{\raggedright\arraybackslash}p{0.34\textwidth}
>{\centering\arraybackslash}p{0.24\textwidth}
>{\centering\arraybackslash}X
}
\toprule

\textbf{Random effects structure}
&
\textbf{Random effects contribution}
&
\textbf{Variance--covariance structure}
\\
\midrule

Subject-level random intercept with the same variance across cohorts
&
$\nu_{0i}$
&
\[
\mathbf D
=
\begin{pmatrix}
\tau_0^2
\end{pmatrix}
\] 
\\

\addlinespace

Subject-level random intercept with cohort-specific variance
&
$\nu_{0i}$
&
\[
\mathbf D_{k(i)}
=
\begin{pmatrix}
\tau_{0,k(i)}^2
\end{pmatrix}
\]
\\

\addlinespace



Random intercepts at both the cohort and subject levels
&
$u_{0,k(i)}+\nu_{0i}$
&
\[
\mathbf D_u
=
\begin{pmatrix}
\omega_0^2
\end{pmatrix},
\qquad
\mathbf D_\nu
=
\begin{pmatrix}
\tau_0^2
\end{pmatrix}
\]
\\

\bottomrule
\end{tabularx}

\begin{tablenotes}
\footnotesize
\item \textit{Notes:}
$a_{ij}$ denotes centered age;
$k(i)$ denotes the cohort k containing subject $i$;
$\boldsymbol{\nu}_i$ denotes subject-level random effects;
and $\boldsymbol{u}_k$ denotes cohort-level random effects.
$\mathbf D$ denotes the subject-level random-effects
variance--covariance matrix;
$\mathbf D_{k(i)}$ denotes a cohort-specific subject-level
random-effects variance--covariance matrix;
and $\mathbf D_u$ and $\mathbf D_\nu$ denote the cohort- and
subject-level random-effects variance--covariance matrices,
respectively.
The common random intercept standard deviation in the simulatio study is $\tau_0=0.8$. 
The cohort-specific random intercept standard deviations are $\tau_{0,1}=0.8$ and $\tau_{0,2}=1.5$ for cohort 1 and cohort 2. 
The cohort-level random intercept standard deviation is $\omega_0=0.2$.
\end{tablenotes}
\end{threeparttable}
\end{table}

\begin{table}[t]
\caption{Residual variance--covariance structures considered in
different DGP settings}
\label{tab:residual_structures}
\centering

\begin{threeparttable}

\normalsize
\renewcommand{\arraystretch}{1.35}
\setlength{\tabcolsep}{5pt}

\begin{tabularx}{\textwidth}{
>{\raggedright\arraybackslash}p{0.43\textwidth}
>{\centering\arraybackslash}X
}
\toprule

\textbf{Residual structure}
&
\textbf{Variance--covariance structure $\mathbf R_i$}
\\
\midrule

Constant residual variance and conditionally independent residuals
within subjects
&
\[
\mathbf R_i
=
\sigma^2\mathbf I_{n_i}
\]
\\

\addlinespace

Study wave-specific residual variances and conditionally independent
residuals within subjects
&
\[
\mathbf R_i
=
\mathbf S_i^2
=
\operatorname{diag}
\left(
\sigma_{g_{i1}}^2,\ldots,
\sigma_{g_{in_i}}^2
\right)
\]
\\

\addlinespace

Constant residual variance with continuous-time AR(1) correlation
within subjects
&
\[
\mathbf R_i
=
\sigma^2\mathbf C_i(\rho)
\]
\\

\addlinespace

Study wave-specific residual variances with continuous-time AR(1)
correlation within subjects
&
\[
\mathbf R_i
=
\mathbf S_i
\mathbf C_i(\rho)
\mathbf S_i,
\]
\\

\bottomrule
\end{tabularx}

\begin{tablenotes}
\footnotesize
\item \textit{Notes:}
The constant residual variance in the simulation study is $\sigma^2 = 0.7^2$.
$g_{ij}$ ($j \in \{1,\ldots,n_i\}$) denotes the study wave
associated with observation $j$ of subject $i$, and $\sigma_{kg}^2$
denotes the residual variance for wave $g$ in cohort $k$. 
In the simulation study, there are 4 study waves within each cohort ($\sigma_{11}^2$, $\sigma_{12}^2$, $\sigma_{13}^2$, $\sigma_{14}^2$, $\sigma_{21}^2$, $\sigma_{22}^2$, $\sigma_{23}^2$, $\sigma_{24}^2$) = ($0.8^2$, $0.4^2$, $0.3^2$, $0.3^2$, $0.5^2$, $0.5^2$, $0.3^2$, $0.3^2$).
$\rho$ is the correlation parameter which indicates how quickly correlation decreases as time interval between measurements increases ($\rho=0.8$ in the simulation study).
\end{tablenotes}

\end{threeparttable}
\end{table}

As illustrated in Figure~\ref{f:simulation_design}, our simulation study was conducted under different data-generating process (DGP) settings, with varying specifications of the random effects and residual structures, as shown in Table~\ref{tab:random_effect_structures} and Table~\ref{tab:residual_structures}. The design features, cohort-specific covariate distributions, and the true fixed-effect function are kept identical across all DGP settings. 
For each DGP setting, 20 longitudinal datasets (each consisting of two cohorts $C1$ and $C2$) were simulated. Cohort $C1$ consists of 8,000 subjects with baseline ages uniformly distributed between 4.5 and 7 years, while Cohort $C2$ consists of 15,000 subjects with baseline ages uniformly distributed between 2 and 10 years. Four measurement waves are scheduled at 0, 3, 6, and 11 years after baseline for $C1$ and at 0, 2, 6, and 13 years after baseline
for $C2$. Within each simulated dataset, three training scenarios were constructed to assess prediction of outcomes after age 14 in the target cohort (C1). The test set consisted of 50\% of C1 observations at age 14 or older and was kept identical across the three training scenarios. In scenario 1, the training set included only observations from C1 before age 14. In scenario 2, the training set additionally included the remaining 50\% of the observations from C1 at age 14 or older (after assigning observations to the test set). In scenario 3, the training set combined observations from C1 before age 14 with observations from the second cohort C2, thereby extending the age range being covered in the training data.

Within each training scenario, we considered baseline methods, linear mixed models (LMMs), and mixed-effects random forest (MERF) approaches. The baseline methods include prediction using the subject's most recent observation or the mean of the subject's previous observations, linear models with correctly specified or mis-specified fixed effects (LM and LM\_misspec, respectively), and random forest (RF). Similarly, for the LMMs, we considered models with correctly specified or mis-specified fixed effects (LMM and LMM\_misspec, respectively).
For MERF, the standard MERF \citep{Hajjem2014} assumed the same random-effects variance--covariance matrix for all subjects, i.e., $\mathbf{D}_i=\mathbf{D}$, and independent residuals with constant variance, i.e., $\boldsymbol{R}_i=\sigma_\varepsilon^2\mathbf{I}_{n_i}$. In contrast, MERF$_{multi-cohort}$ allowed the random effects and residual variance--covariance structures to correspond to the true DGP. 

We first evaluated these models under the DGP setting with the simplest variance-covariance structure, consisting of subject-level
random intercepts with a common variance across cohorts and
homoscedastic independent residuals. This setting is compatible with
the covariance assumptions of standard MERF \citep{Hajjem2014} and therefore provides a benchmark for assessing whether the additional flexibility of MERF$_{multi-cohort}$ affects performance when a
more complex variance--covariance structure is not required. We then
considered DGP settings with increasingly complex variance-covariance structures to assess whether allowing more flexible random effect and residual structures improves prediction performance. 

Prediction performance was evaluated by the mean absolute error (MAE) on the same test set within each simulation iteration. For each method and training scenario, we report the mean of the test-set MAE across 20 simulation iterations together with its Monte Carlo standard error (MCSE). 
Since the true covariance parameters $\boldsymbol\psi$ are known under the simulation
DGP, we additionally examined recovery of random effects (co)variance  and residual variance. Recovery of covariance parameters are evaluated using the mean of the parameter estimates across 20 simulation iterations together with its Monte Carlo standard error (MCSE).


\subsection{Simulation Results}
\label{sec:simulation_results}

\subsubsection{Under the simple DGP setting}
\label{sec:simple_dgp_results}


The simple DGP provides a setting in which the standard MERF covariance assumptions are met. Table~\ref{tab:simple_prediction} reports the mean and SD of the test-set MAE, together with the corresponding Monte Carlo standard error (MCSE) in 20 simulation iterations, for all candidate methods across three training scenarios. Table~\ref{tab:simple_parameter_recovery} summarizes parameter recovery of the subject-level
random intercept standard deviation $\tau_0$ and residual variance $\sigma^2$ for the models incorporating random effects.

\begin{table}[t]
\centering
\caption{Prediction performance under the simple DGP across the three training scenarios}
\label{tab:simple_prediction}

\normalsize
\renewcommand{\arraystretch}{1.15}
\setlength{\tabcolsep}{8pt}

\begin{tabularx}{\textwidth}{
>{\raggedright\arraybackslash}X
>{\centering\arraybackslash}p{0.20\textwidth}
>{\centering\arraybackslash}p{0.20\textwidth}
>{\centering\arraybackslash}p{0.20\textwidth}
}
\toprule

\textbf{Method}
&
\textbf{Scenario 1}
&
\textbf{Scenario 2}
&
\textbf{Scenario 3}
\\
&
\textbf{mean (MCSE)}
&
\textbf{mean (MCSE)}
&
\textbf{mean (MCSE)}
\\

\midrule

Mean
& 1.068 (0.006) & 1.068 (0.006) & 1.068 (0.006) \\
Previous
& 1.323 (0.007) & 1.323 (0.007) & 1.323 (0.007) \\

LM
& 0.911 (0.003) & 0.910 (0.003) & 0.909 (0.003) \\

LM$_{\text{misspec}}$
&0.989 (0.003) & 0.981 (0.003) & 0.977 (0.003) \\

RF
&0.933 (0.004) &0.917 (0.003) &0.911 (0.003) \\

LMM
&0.647 (0.003) &0.645 (0.003) & 0.645 (0.003) \\

LMM$_{\text{misspec}}$
&0.764 (0.004) &0.754 (0.004) & 0.750 (0.004) \\

MERF
& 0.691 (0.004) & 0.660 0.003)& 0.650 (0.003)\\

MERF$_{\text{multi-cohort}}$
& 0.696 (0.004)& 0.663 (0.003)& 0.651 (0.002)\\

\bottomrule
\end{tabularx}

\begin{tablenotes}
\footnotesize
\item \textit{Notes:}
mean denotes the mean test-set mean absolute error (MAE) across 20 simulation
iterations, with Monte Carlo standard errors (MCSE) in parentheses.
\end{tablenotes}
\end{table}

Under the simple DGP, MERF$_{multi-cohort}$ shows predictive performance comparable to that of standard MERF across all three training scenarios. Within each training scanario, the differences in mean test-set MAEs between  MERF$_{multi-cohort}$ and MERF are small relative to the between-repetition variation and decreased as the amount of training information increased. 

A clearer difference is observed in the recovery of the covariance parameters. Standard MERF systematically underestimates the subject-level random-intercept standard deviation compared with the true value. In contrast, MERF$_{multi-cohort}$ provides estimates closer to the truth. Similarly, standard MERF overestimates the residual variance across the three scenarios, whereas the corresponding estimates from MERF$_{multi-cohort}$ are  consistently closer to the true value. These results suggest that the proposed estimation procedure with subject-level cross-fitting and affine calibration of fixed effect improves the separation between the fixed effect component and the stochastic components, while retaining predictive performance comparable to standard MERF.

The correctly specified LMM provides the best predictive performance and the most accurate recovery of the covariance parameters. This is expected because the functional form of the fixed effects used to generate the data is explicitly specified as fixed effects in LMM. However, the true fixed-effects function which may contain complex nonlinearities and interactions are unknown in practice. This limitation is illustrated by LMM$_{misspec}$, which shows larger prediction error than MERF$_{multi-cohort}$ across all three training scenarios. Although LMM$_{misspec}$ recovers the random intercept standard deviation well, it overestimates the residual variance across three scenarios. This shows that mis specification of the fixed effect component can leave population-level variation unexplained, which may subsequently be attributed to the residual component.

\begin{table}[t]
\centering
\caption{Recovery of variance components under the simple DGP}
\label{tab:simple_parameter_recovery}

\normalsize
\renewcommand{\arraystretch}{1.15}
\setlength{\tabcolsep}{4pt}

\begin{tabularx}{\textwidth}{
>{\raggedright\arraybackslash}p{0.11\textwidth}
>{\centering\arraybackslash}p{0.08\textwidth}
>{\raggedright\arraybackslash}p{0.18\textwidth}
*{3}{>{\centering\arraybackslash}X}
}
\toprule

\textbf{Parameter}
&
\textbf{Truth}
&
\textbf{Method}
&
\textbf{Scenario 1}
&
\textbf{Scenario 2}
&
\textbf{Scenario 3}
\\

&
&
&
\textbf{mean (MCSE)}
&
\textbf{mean (MCSE)}
&
\textbf{mean (MCSE)}
\\
\midrule

$\tau_0$
&
0.800
&
LMM
&
0.800 (0.004)
&
0.800 (0.004)
&
0.800 (0.004)
\\

&
&
LMM$_{\text{misspec}}$
&
0.801 (0.004)
&
0.801 (0.003)
&
0.799 (0.002)
\\

&
&
MERF
&
0.756 (0.004)
&
0.769 (0.005)
&
0.776 (0.003)
\\

&
&
MERF$_{\text{multi-cohort}}$
&
0.801 (0.004)
&
0.802 (0.004)
&
0.804 (0.002)
\\

\addlinespace

$\sigma^2$
&
0.490
&
LMM
&
0.495 (0.002)
&
0.496 (0.002)
&
0.495 (0.001)
\\

&
&
LMM$_{\text{misspec}}$
&
0.709 (0.002)
&
0.711 (0.002)
&
0.822 (0.002)
\\

&
&
MERF
&
0.554 (0.003)
&
0.544 (0.004)
&
0.532 (0.002)
\\

&
&
MERF$_{\text{multi-cohort}}$
&
0.522 (0.002)
&
0.522 (0.002)
&
0.520 (0.001)
\\

\bottomrule
\end{tabularx}
\begin{tablenotes}
\footnotesize
\item \textit{Notes:}
mean and MCSE denote the mean of the parameter estimates across 20 simulation iterations and corresponding Monte Carlo standard error.

\end{tablenotes}
\end{table}

\subsubsection{Under more complex DGP settings}
\label{sec:complex_dgp_results}

We next evaluate the methods under DGP settings with increasingly complex between- and within-subject covariance structures, including cohort-specific random-effects variances, cohort-level random effects, study wave-specific
residual variances, and continuous-time AR(1) residual correlation.
These settings were used to assess whether prediction performance and estimation of the covariance parameters can be improved by MERF$_{multi-cohort}$ when the covariance assumptions in standard MERF are violated.

Under the simple DGP, standard MERF \citep{Hajjem2014} was
implemented using the \texttt{LongituRF} package \citep{LongituRF}. The improved recovery of the covariance parameters by MERF$_{multi-cohort}$ relative to standard MERF \citep{Hajjem2014} is consistent with the aim of the proposed subject-level cross-fitting and affine calibration procedures. 
Under the more complex DGP settings, to assess whether allowing more flexible variance-covariance structures improves covariance parameter recovery, we compared MERF$_{multi-cohort}$ with a MERF fitted using the same proposed estimation procedure, including subject-level cross-fitting and affine calibration, but retaining the same random effects and residual covariance structures as standard MERF. Thus, under more complex DGP settings, MERF$_{multi-cohort}$ and MERF differed only in their variance-covariance structure specification, allowing the effect of flexibility in the specification of the random effects
structure and residual variance--covariance structure to be assessed.

\begin{table}[t]
\centering
\caption{Prediction performance under the more complex DGP settings
across the three training scenarios}
\label{tab:complex_prediction}

\normalsize
\renewcommand{\arraystretch}{1.15}
\setlength{\tabcolsep}{4pt}

\begin{tabularx}{\textwidth}{
>{\raggedright\arraybackslash}p{0.25\textwidth}
>{\raggedright\arraybackslash}p{0.16\textwidth}
>{\centering\arraybackslash}X
>{\centering\arraybackslash}X
>{\centering\arraybackslash}X
}
\toprule

\textbf{DGP setting}
&
\textbf{Method}
&
\textbf{Scenario 1}
&
\textbf{Scenario 2}
&
\textbf{Scenario 3}
\\

&
&
\textbf{mean (MCSE)}
&
\textbf{mean (MCSE)}
&
\textbf{mean (MCSE)}
\\

\midrule


\multirow[t]{9}{=}{
Random intercepts at both the cohort and subject levels;
constant residual variance and conditionally independent
residuals within subjects
}
&
Mean
& 1.057 (0.004) & 1.057 (0.004) & 1.057 (0.004) \\

&
Previous
& 1.306 (0.007) & 1.306 (0.007) & 1.306 (0.007) \\

&
LM
& 0.912 (0.005) & 0.911 (0.005) & 0.971 0.018) \\

&
LM$_{\text{misspec}}$
& 0.983 (0.006) & 0.975 (0.005) & 1.056 (0.024) \\

&
RF
& 0.933 (0.004) & 0.913 (0.005) & 1.031 ( 0.035) \\

&
LMM
& \text{only one cohort} & \text{only one cohort} & 0.640 (0.003) \\

&
LMM$_{\text{misspec}}$
& \text{only one cohort} & \text{only one cohort} & 0.742 (0.004) \\

&
MERF
& 0.685 (0.003) & 0.658 (0.003)& 0.651 (0.003) \\

&
MERF$_{\text{multi-cohort}}$
& \text{only one cohort} & \text{only one cohort} & 0.645 (0.003) \\

\midrule


\multirow[t]{9}{=}{
Subject-level random intercepts with cohort-specific variance;
constant residual variance and conditionally independent
residuals within subjects
}
&
Mean
& 1.051 (0.008) & 1.051 (0.008) & 1.051 (0.008) \\

&
Previous
& 1.310 (0.008) & 1.310 (0.008) & 1.310 (0.008) \\

&
LM
& 0.911 (0.004) & 0.910 (0.004) & 0.910 (0.004) \\

&
LM$_{\text{misspec}}$
& 0.987 (0.004) & 0.978 (0.004) & 0.975 (0.004) \\

&
RF
& 0.934 (0.005) & 0.916 (0.005) & 0.914 (0.005) \\

&
LMM
& \text{only one cohort} & \text{only one cohort} & 0.636 (0.003) \\

&
LMM$_{\text{misspec}}$
& \text{only one cohort} & \text{only one cohort} & 0.742 (0.004) \\

&
MERF
& 0.690 (0.005) & 0.658 (0.004) & 0.647 (0.003) \\

&
MERF$_{\text{multi-cohort}}$
& \text{only one cohort} & \text{only one cohort} & 0.646 (0.003) \\

\midrule


\multirow[t]{9}{=}{
Subject-level random intercepts;
study wave-specific residual variances and conditionally
independent residuals within subjects
}
&
Mean
& 0.887 (0.005) & 0.887 (0.005)  & 0.887 (0.005) \\

&
Previous
& 1.070 (0.007) & 1.070 (0.007)  & 1.070 (0.007) \\

&
LM
& 0.764 (0.003) & 0.763 (0.003) & 0.762 (0.003) \\

&
LM$_{\text{misspec}}$
& 0.848 (0.004) & 0.837 (0.004) & 0.834 (0.004) \\

&
RF
& 0.778 (0.004) & 0.763 (0.004) & 0.760 (0.004) \\

&
LMM
& 0.328 (0.002) & 0.323 (0.002) & 0.326 (0.002) \\

&
LMM$_{\text{misspec}}$
& 0.530 (0.003) & 0.500 (0.003) & 0.493 (0.003) \\
&
MERF
& 0.426 (0.004) & 0.381 (0.002)  & 0.363 (0.002) \\

&
MERF$_{\text{multi-cohort}}$
& 0.403 (0.004) & 0.355 (0.002)  & 0.333 (0.002) \\

\midrule


\multirow[t]{9}{=}{
Subject-level random intercepts;
constant residual variance with continuous-time AR(1)
correlation within subjects
}
&
Mean
& 1.068 (0.004) & 1.068 (0.004) & 1.068 (0.004) \\

&
Previous
& 1.316 (0.006) & 1.316 (0.006) & 1.316 (0.006) \\

&
LM
& 0.916 (0.005) & 0.915 (0.005) & 0.914 (0.005) \\

&
LM$_{\text{misspec}}$
& 0.994 (0.005) & 0.985 (0.005) & 0.983 (0.005) \\

&
RF
& 0.940 (0.005) & 0.920 (0.005) & 0.915 (0.005) \\

&
LMM
& 0.650 (0.003) & 0.649 (0.003) & 0.648 (0.003) \\

&
LMM$_{\text{misspec}}$
& 0.769 (0.003) & 0.760 (0.003) & 0.754 (0.003) \\

&
MERF
& 0.702 (0.005) & 0.669 (0.002) & 0.659 (0.003) \\

&
MERF$_{\text{multi-cohort}}$
& 0.700 (0.005) & 0.668 (0.003) & 0.658 (0.003) \\

\bottomrule
\end{tabularx}

\begin{tablenotes}
\footnotesize
\item \textit{Notes:}
values are the mean test-set mean absolute error (MAE) across
20 simulation repetitions, with the Monte Carlo standard error (MCSE) in parentheses.
Cohort level random effects and cohort-specific random effect variances could only be estimated with traning data from 2 cohorts.
\end{tablenotes}

\end{table}

\begin{table}[t]
\centering
\caption{Recovery of variance-covariance parameters under complex
random effects DGP settings across the three training scenarios}
\label{tab:complex_re_recovery}

\normalsize
\renewcommand{\arraystretch}{1.15}
\setlength{\tabcolsep}{3pt}

\begin{tabularx}{\textwidth}{
>{\raggedright\arraybackslash}p{0.23\textwidth}
>{\centering\arraybackslash}p{0.08\textwidth}
>{\centering\arraybackslash}p{0.06\textwidth}
>{\raggedright\arraybackslash}p{0.15\textwidth}
>{\centering\arraybackslash}X
>{\centering\arraybackslash}X
>{\centering\arraybackslash}X
}
\toprule

\textbf{DGP setting}
&
\textbf{Parameter}
&
\textbf{Truth}
&
\textbf{Method}
&
\textbf{Scenario 1}
&
\textbf{Scenario 2}
&
\textbf{Scenario 3}
\\

&
&
&
&
\textbf{mean (MCSE)}
&
\textbf{mean (MCSE)}
&
\textbf{mean (MCSE)}
\\
\midrule


\multirow[t]{12}{=}{
Random intercepts at both the cohort and subject levels;
constant residual variance and conditionally independent
residuals within subjects
}
&
\multirow{4}{*}{$\omega_0$}
&
\multirow{4}{*}{0.200}
&
LMM & \text{only one cohort} & \text{only one cohort} & 0.105 (0.037) \\

& & & LMM$_{\text{misspec}}$ & \text{only one cohort} &\text{only one cohort} & 0.106 (0.035) \\
& & & MERF & -- & -- & -- \\
& & & MERF$_{\text{multi-cohort}}$ & \text{only one cohort}  & \text{only one cohort} &0.104 (0.035) \\

\addlinespace

&
\multirow{4}{*}{$\tau_0$}
&
\multirow{4}{*}{0.800}
&
LMM & \text{only one cohort} & \text{only one cohort} & 0.800 (0.002) \\

& & & LMM$_{\text{misspec}}$ & \text{only one cohort}&\text{only one cohort} & 0.791 (0.003) \\
& & & MERF & 0.801 (0.005) & 0.799 (0.005) & 0.885 (0.003) \\
& & & MERF$_{\text{multi-cohort}}$ & \text{only one cohort}  & \text{only one cohort} &0.801 (0.002) \\

\addlinespace

&
\multirow{4}{*}{$\sigma^2$}
&
\multirow{4}{*}{0.490}
&

LMM & \text{only one cohort} & \text{only one cohort} & 0.495 (0.001) \\

& & & LMM$_{\text{misspec}}$ & \text{only one cohort}&\text{only one cohort}& 0.825 (0.002) \\
& & & MERF & 0.522 (0.002) & 0.523 (0.002) & 0.526 (0.001) \\
& & & MERF$_{\text{multi-cohort}}$ & \text{only one cohort}  & \text{only one cohort} &0.526 (0.001) \\

\midrule


\multirow[t]{12}{=}{
Subject-level random intercepts with cohort-specific variance;
constant residual variance and conditionally independent
residuals within subjects
}
&
\multirow{4}{*}{$\tau_{0,1}$}
&
\multirow{4}{*}{0.800}
&
LMM & \text{only one cohort} & \text{only one cohort} & 0.804 (0.004) \\

& & & LMM$_{\text{misspec}}$ &\text{only one cohort} &\text{only one cohort} & 0.755 (0.005) \\
& & & MERF & 0.806 (0.005) & 0.806 (0.005) & 1.260 (0.003) \\
& & & MERF$_{\text{multi-cohort}}$ & \text{only one cohort} & \text{only one cohort} & 0.799 (0.004) \\

\addlinespace

&
\multirow{4}{*}{$\tau_{0,2}$}
&
\multirow{4}{*}{1.500}
&
LMM & \text{only one cohort} & \text{only one cohort} & 1.500 (0.004) \\

& & & LMM$_{\text{misspec}}$ & \text{only one cohort} & \text{only one cohort} & 1.510 (0.004) \\
& & & MERF & 0.806 (0.005) & 0.806 (0.005) & 1.260 (0.003) \\
& & & MERF$_{\text{multi-cohort}}$ &\text{only one cohort} & \text{only one cohort} & 1.510 (0.004) \\

\addlinespace

&
\multirow{4}{*}{$\sigma^2$}
&
\multirow{4}{*}{0.490}
&
LMM & \text{only one cohort} &\text{only one cohort} & 0.495 (0.001) \\

& & & LMM$_{\text{misspec}}$ & \text{only one cohort} & \text{only one cohort} & 0.825 (0.002) \\
& & & MERF & 0.522 (0.002) & 0.523 (0.002) & 0.524 (0.001) \\
& & & MERF$_{\text{multi-cohort}}$ & \text{only one cohort} & \text{only one cohort} & 0.525 (0.001) \\

\bottomrule
\end{tabularx}

\begin{tablenotes}
\footnotesize
\item \textit{Notes:}
values are the mean parameter estimates across 20 simulation
repetitions, with Monte Carlo standard error (MCSE) in parentheses.
$\omega_0$ denotes the cohort-level random-intercept standard deviation.
$\tau_{0,k}$ and $\tau_{1,k}$ denote the subject-level random-intercept
and random-slope standard deviations, respectively, in cohort $k$.
$\tau_{01,k}$ denotes the corresponding intercept--slope covariance.
and $\sigma^2$ denotes the residual variance.
Entries are reported as "--" when the cohort-level random intercept standard deviation parameter is not estimated by the MERF because MERF does not support nested random effects.
\end{tablenotes}

\end{table}

\begin{table}[t]
\centering
\caption{Recovery of variance--covariance parameters under complex
residual DGP settings across the three training scenarios}
\label{tab:complex_residual_recovery}

\normalsize
\renewcommand{\arraystretch}{1}
\setlength{\tabcolsep}{3pt}

\begin{tabularx}{\textwidth}{
>{\raggedright\arraybackslash}p{0.23\textwidth}
>{\centering\arraybackslash}p{0.08\textwidth}
>{\centering\arraybackslash}p{0.06\textwidth}
>{\raggedright\arraybackslash}p{0.15\textwidth}
>{\centering\arraybackslash}X
>{\centering\arraybackslash}X
>{\centering\arraybackslash}X
}
\toprule

\textbf{DGP setting}
&
\textbf{Parameter}
&
\textbf{Truth}
&
\textbf{Method}
&
\textbf{Scenario 1}
&
\textbf{Scenario 2}
&
\textbf{Scenario 3}
\\

&
&
&
&
\textbf{mean (MCSE)}
&
\textbf{mean (MCSE)}
&
\textbf{mean (MCSE)}
\\
\midrule


\multirow[t]{12}{=}{
Subject-level random intercepts;
study wave-specific residual variances and conditionally
independent residuals within subjects
}
&
\multirow{4}{*}{$\tau_0$}
&
\multirow{4}{*}{0.800}
&
LMM & 0.800 (0.004) & 0.800 (0.004) & 0.798 (0.002) \\

& & & LMM$_{\text{misspec}}$ & 0.799 (0.004) & 0.800 (0.004) & 0.801 (0.002) \\
& & & MERF & 0.881 (0.005) & 0.891 (0.005) & 0.851 (0.002) \\
& & & MERF$_{\text{multi-cohort}}$ &0.809 (0.004) & 0.811 (0.004) & 0.810 (0.002) \\

\addlinespace

&
\multirow{4}{*}{$\sigma_{11}^2$}
&
\multirow{4}{*}{0.640}
&
LMM
& 0.641 (0.003)
& 0.641 (0.003)
& 0.642 (0.003)
\\

& & &
LMM$_{\text{misspec}}$
& 0.852 (0.006)
& 0.849 (0.005)
& 0.874 (0.006)
\\

& & &
MERF
&0.351 (0.007) &0.328 (0.006) &0.260 (0.003)
\\

& & &
MERF$_{\text{multi-cohort}}$
& 0.667 (0.003)
& 0.668 (0.003)
& 0.649 (0.004)
\\

\addlinespace

&
\multirow{4}{*}{$\sigma_{12}^2$}
&
\multirow{4}{*}{0.160}
&
LMM
& 0.160 (0.001)
& 0.160 (0.001)
& 0.160 (0.001)
\\

& & &
LMM$_{\text{misspec}}$
& 0.371 (0.004)
& 0.367 (0.004)
& 0.391 (0.004)
\\

& & &
MERF
&0.351 (0.007) &0.328 (0.006) &0.260 (0.003)
\\

& & &
MERF$_{\text{multi-cohort}}$
& 0.191 (0.001)
& 0.189 (0.001)
& 0.175 (0.001)
\\

\addlinespace

&
\multirow{4}{*}{$\sigma_{13}^2$}
&
\multirow{4}{*}{0.090}
&
LMM
& 0.092 (0.001)
& 0.093 (0.001)
& 0.095 (0.001)
\\

& & &
LMM$_{\text{misspec}}$
& 0.317 (0.003)
& 0.330 (0.003)
& 0.326 (0.004)
\\

& & &
MERF
&0.351 (0.007) &0.328 (0.006) &0.260 (0.003)
\\

& & &
MERF$_{\text{multi-cohort}}$
& 0.143 (0.002)
& 0.135 (0.002)
& 0.110 (0.001)
\\

\addlinespace

&
\multirow{4}{*}{$\sigma_{14}^2$}
&
\multirow{4}{*}{0.090}
&
LMM
&--
& 0.111 (0.002)
&--
\\

& & &
LMM$_{\text{misspec}}$
&--
& 0.321 (0.006)
&--
\\

& & &
MERF
&-- &0.328 (0.006) &--
\\

& & &
MERF$_{\text{multi-cohort}}$
&--
& 0.141 (0.003)
&--
\\

\addlinespace

&
\multirow{4}{*}{$\sigma_{21}^2$}
&
\multirow{4}{*}{0.250}
&
LMM
&--
&--
& 0.248 (0.001)
\\

& & &
LMM$_{\text{misspec}}$
&--
&--
& 0.589 (0.003)
\\

& & &
MERF
&-- &-- &0.260 (0.003)
\\

& & &
MERF$_{\text{multi-cohort}}$
&--
&--
& 0.285 (0.002)
\\

\addlinespace

&
\multirow{4}{*}{$\sigma_{22}^2$}
&
\multirow{4}{*}{0.250}
&
LMM
&--
&--
& 0.254 (0.001)
\\

& & &
LMM$_{\text{misspec}}$
&--
&--
& 0.591 (0.003)
\\

& & &
MERF
&-- &-- &0.260 (0.003)
\\

& & &
MERF$_{\text{multi-cohort}}$
&--
&--
& 0.285 (0.002)
\\

\addlinespace

&
\multirow{4}{*}{$\sigma_{23}^2$}
&
\multirow{4}{*}{0.090}
&
LMM
&--
&--
& 0.107 (0.001)
\\

& & &
LMM$_{\text{misspec}}$
&--
&--
& 0.529 (0.004)
\\

& & &
MERF
&-- &--&0.260 (0.003)
\\

& & &
MERF$_{\text{multi-cohort}}$
&--
&--
& 0.150 (0.002)
\\

\addlinespace

&
\multirow{4}{*}{$\sigma_{24}^2$}
&
\multirow{4}{*}{0.090}
&
LMM
&--
&--
& 0.099 (0.001)
\\

& & &
LMM$_{\text{misspec}}$
&--
&--
& 0.503 (0.004)
\\

& & &
MERF
&-- &-- &0.260 (0.003)
\\

& & &
MERF$_{\text{multi-cohort}}$
&--
&--
& 0.152 (0.002)
\\

\midrule

\multirow[t]{8}{=}{
Subject-level random intercepts;
constant residual variance with continuous-time AR(1)
correlation within subjects
}
&
\multirow{4}{*}{$\tau_0$}
&
\multirow{4}{*}{0.800}
&
LMM & 0.808 (0.004) & 0.798 (0.004) & 0.802 (0.002) \\

& & & LMM$_{\text{misspec}}$ & 0.805 (0.005) & 0.800 (0.004) & 0.789 (0.004) \\
& & & MERF & 0.855 (0.004) & 0.847 (0.004) & 0.857 (0.003) \\
& & & MERF$_{\text{multi-cohort}}$ & 0.807 (0.004) & 0.801 (0.004) & 0.805 (0.003) \\

\addlinespace

&
\multirow{4}{*}{$\sigma^2$}
&
\multirow{4}{*}{0.490}
&
LMM & 0.486 (0.002) & 0.496 (0.002) & 0.496 (0.001) \\

& & & LMM$_{\text{misspec}}$ & 0.701 (0.005) & 0.707 (0.003) & 0.836 (0.002) \\
& & & MERF& 0.474 (0.001) & 0.483 (0.001) & 0.478 (0.001) \\
& & & MERF$_{\text{multi-cohort}}$ & 0.501 (0.003) & 0.506 (0.002) & 0.506 (0.001) \\

\bottomrule
\end{tabularx}

\begin{tablenotes}
\footnotesize
\item \textit{Notes:}
values are the mean parameter estimates across 20 simulation
repetitions, with Monte Carlo standard error (MCSE) in parentheses.
$\tau_0$ denotes the subject-level random-intercept standard deviation.
$\sigma^2$ denotes the common residual variance. $\sigma_{kg}^2$ denotes the study wave-specific residual variance for wave $g$ in cohort $k$. Entries are reported as "--" when the study wave-specific residual variance parameter is not estimated by the fitted model because training data do not include observation from wave $g$ in cohort $k$. 
\end{tablenotes}

\end{table}

Prediction performance under the more complex DGP settings is summarized in Table~\ref{tab:complex_prediction}.
Table~\ref{tab:complex_re_recovery} and Table~\ref{tab:complex_residual_recovery} summarize the recovery of variance-covariance parameters across the three training scenarios under different DGP settings with complex random effects structures and residual structures respectively. 

Across the more complex DGP settings, the main advantage of MERF$_{multi-cohort}$ over MERF is the improvement in estimating the variance-covariance component rather than lower prediction error. 

When random intercepts at both the cohort and subject levels are present, the comparison between MERF$_{multi-cohort}$ and MERF is available in Scenario 3 where the training data contains more than one cohort which allows the hierarchical structure to be estimated. Including a random effect for cluster membership is usually not recommended when there are very few clusters (e.g., fewer than five as a rule of thumb \citep{Bolker2009,Bolker2015,Gomes2022}) because the between-cluster variance is estimated from too little information, often leading to imprecise variance component estimates. In our setting with two-cohort data, MERF$_{multi-cohort}$ recovers the subject-level random intercept variance as well as the correctly specified LMM. The cohort-level random intercept variance is poorly recovered in all models. This may also explain why incorporating cohort-level random-effect predictions resulted in no obvious improvement in predictive performance.

When the subject-level random intercept variability differed between cohorts, the advantage of the more flexible covariance specification is more apparent in estimating random effect variances than in improving prediction performance. In Scenario 3, MERF and MERF$_{multi-cohort}$ achieved nearly identical predictive performance. MERF$_{multi-cohort}$, which allows the subject-level random-intercept variance to differ by cohort, provides the random intercept variance close to the true value. In contrast, MERF imposes a single common random-intercept variance and therefore removes information about how between-subject heterogeneity differs between cohorts.

The benefit of the more flexible covariance specification is more obvious when residual variances differed across study waves. This setting is relevant to multi-cohort data analyses because outcome measurements may differ across assay methods and laboratories, potentially introducing measurement heterogeneity across measurement occasions. In this setting, MERF$_{multi-cohort}$ consistently achieves better predictive performance than MERF and provides estimates of the subject-level random intercept variance that are closer to the true value. When residual heteroscedasticity is ignored, MERF overestimates random intercept variance in all three scenarios. This indicates that part of the wave-specific residual variation may instead be attributed to between-subject heterogeneity.

Under continuous-time AR(1) residual correlation, MERF$_{multi-cohort}$ and  MERF show similar predictive performance, but MERF$_{multi-cohort}$ provides estimates of the subject-level random intercept variance and residual variance closer to the true value. MERF assumes conditionally independent residuals and therefore does not consider the additional within-subject serial residual correlation. MERF overestimates the random intercept variance across scenarios, which suggests part of the unmodeled residual dependence being attributed to the random effect component. 

Compared with the mis-specified LMM, MERF$_{multi-cohort}$ consistently achieves better prediction performance in all complex DGP settings. Additionally, the misspecified LMM tends to overestimate the residual variance, suggesting that population-level variation not captured by the fixed effects is partly retained in the stochastic component of the model. In contrast, MERF$_{multi-cohort}$ allows the fixed effect function to be estimated flexibly without requiring nonlinearities and interactions to be specified in advance. This flexibility is particularly relevant in real-data applications, where the functional form of the fixed effect is generally unknown. 

Although pooling cohorts increases sample size and extends age coverage, the additional cohort may also introduce heterogeneity in the population distribution and variance-covariance structure. When such heterogeneity is not adequately represented by the fitted model, the additional data may provide little benefit and, in some settings, result in poorer prediction performance. Under the DGP containing both cohort- and subject-level random intercepts, where the prediction errors of LM, LM$_{misspec}$, and RF increases in Scenario 3 after training data in Scenario 1 further incorporates data from the additional cohort. MERF$_{multi-cohort}$ can use the additional cohort while explicitly modeling the hierarchical structure in multi-cohort data. Under the DGP with both cohort- and subject-level random intercepts, MERF$_{multi-cohort}$ achieves a mean MAE of 0.645 in Scenario 3, compared with 0.685 for MERF in Scenario 1. Similar patterns are observed under the other complex DGP settings, with MERF$_{multi-cohort}$ in Scenario 3 achieving better prediction performance than models trained under Scenario 1 and achieving similar performance as models trained under Scenario 2. MERF$_{multi-cohort}$ provides a framework for incorporating  information from other cohort and extend observation period while allowing the random effects and residual variance-covariance structures to accommodate the heterogeneity introduced by the additional cohort.

\section{Application to Real Data}
\label{s:setup}
Harmonized data of two children cohorts will be used as a real-data example to illustrate our approach. We will predict age- and sex-specific BMI z-scores \citep{IOTF} based on comprehensive information on demographics, lifestyle, and early-life factors. A detailed description of the data sources and preprocessing are provided in the online Supplementary Material A.

\begin{figure}[t]
  \centering
  \includegraphics[width=\textwidth]{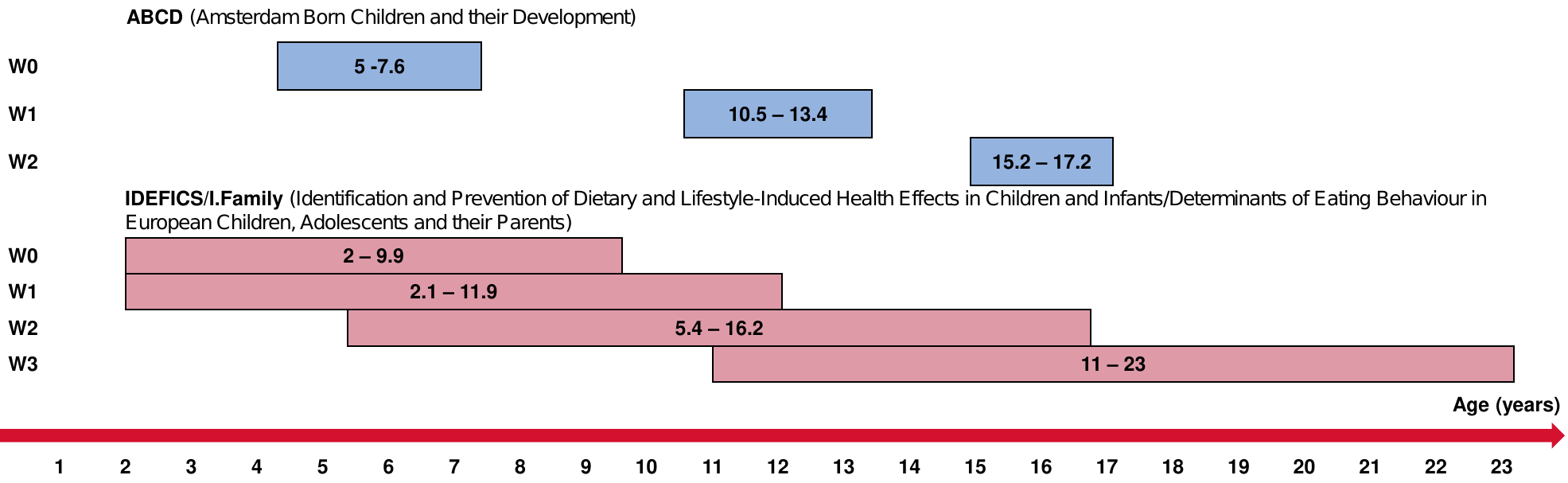}
  \caption{Age coverage in IDEFICS/I.Family cohort and ABCD cohort}
  \label{f:twoCohort}
\end{figure}

The target cohort in our example is the ABCD \citep{abcd} cohort, which consists of 8266 children residing in Amsterdam who were followed up since pregnancy. Three follow-ups (W0, W1 and W2) took place at around the ages of 6 years, 12 years and 16 years, respectively. 
As second cohort we consider the multi-center population-based IDEFICS/I.Family cohort \citep{idefics} covering eight European countries (Belgium, Cyprus, Estonia, Germany, Hungary, Italy, Spain, and Sweden). Compared with ABCD, the IDEFICS/I.Family study includes a larger sample (16229 children at the baseline wave W0) drawn from a more heterogeneous population. As shown in Figure~\ref{f:twoCohort}, the age coverage in the ABCD study includes gaps, whereas IDEFICS/I.Family spans a broader age range. In the IDEFICS/I.Family cohort, the baseline examination wave (W0) included children aged 2.0--9.9 years. Three subsequent follow-up waves were conducted approximately 2 (W1), 6 (W2), and 13 (W3) years after baseline spanning data from childhood to early adulthood. These different cohort designs enable us to investigate the benefits of using multi-cohort rather than single-cohort data to enhance the accuracy of individual-level predictions, particularly in cases where single cohorts are limited in the age range covered.

\begin{figure}[t]
  \centering

  \includegraphics[width=0.70\linewidth,
  height=0.2\textheight,
  keepaspectratio]{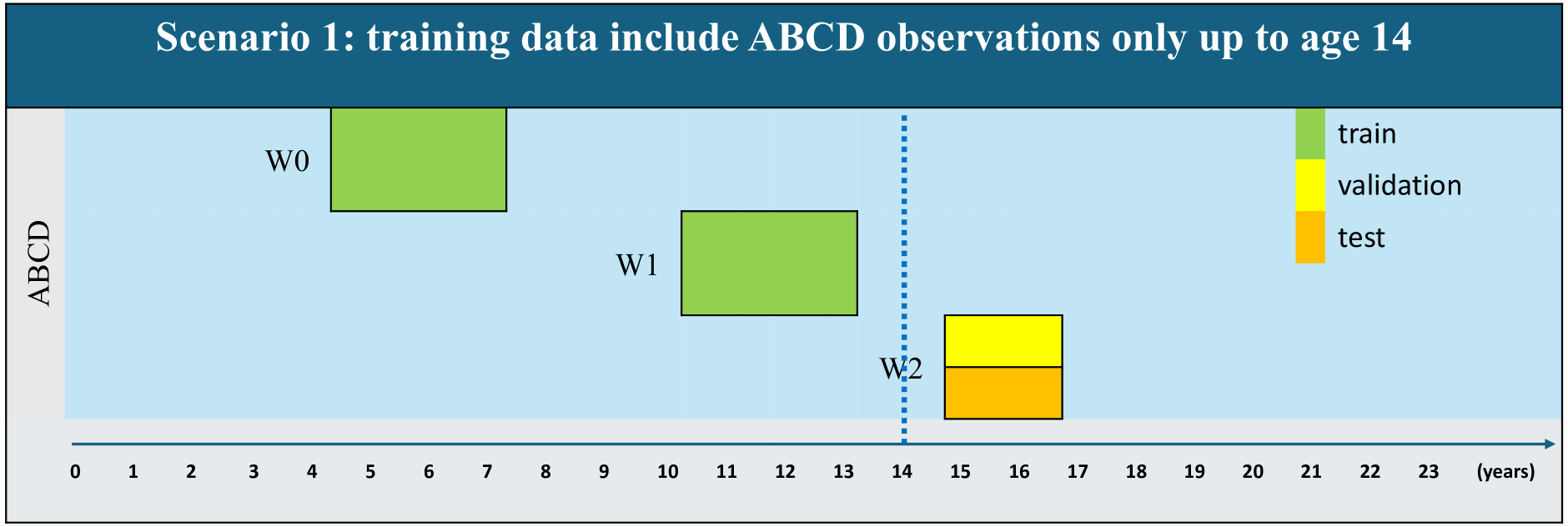}

  \vspace{0.5em}

  \includegraphics[width=0.70\linewidth,
  height=0.2\textheight,
  keepaspectratio]{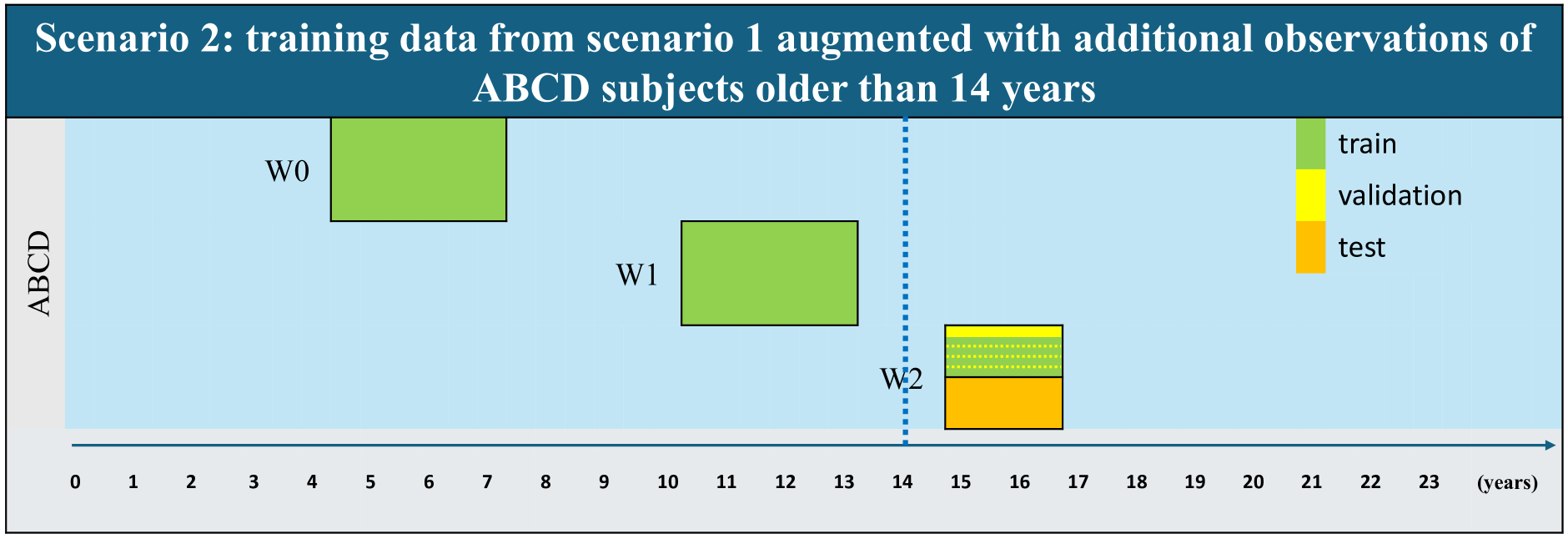}

  \caption{Single-cohort scenarios}
  \label{f:design12}
\end{figure}

\begin{figure}[t]
  \centering

  \includegraphics[width=0.7\linewidth,
  height=0.30\textheight,
  keepaspectratio]{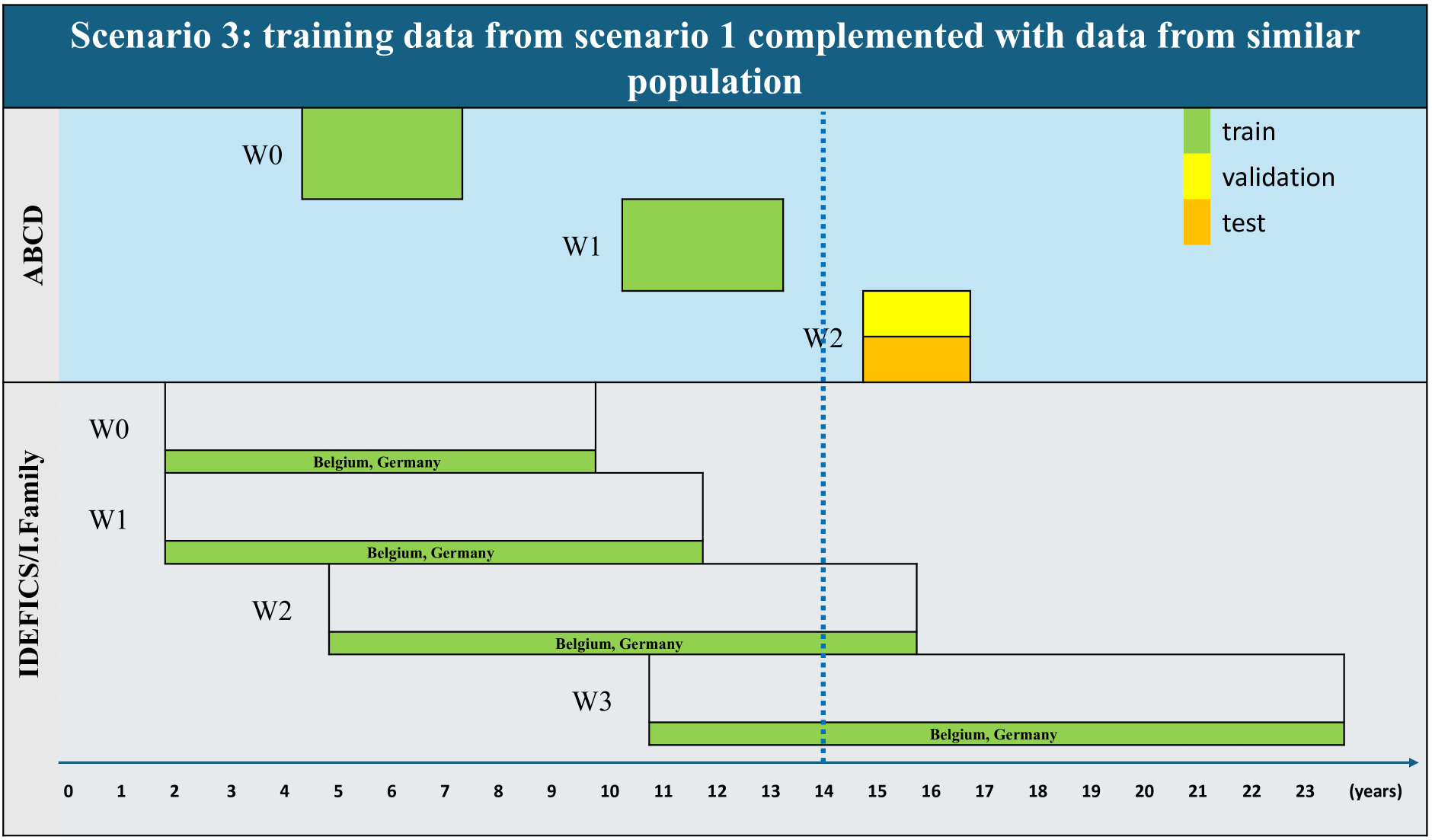}

  \vspace{0.5em}

  \includegraphics[width=0.7\linewidth,
  height=0.30\textheight,
  keepaspectratio]{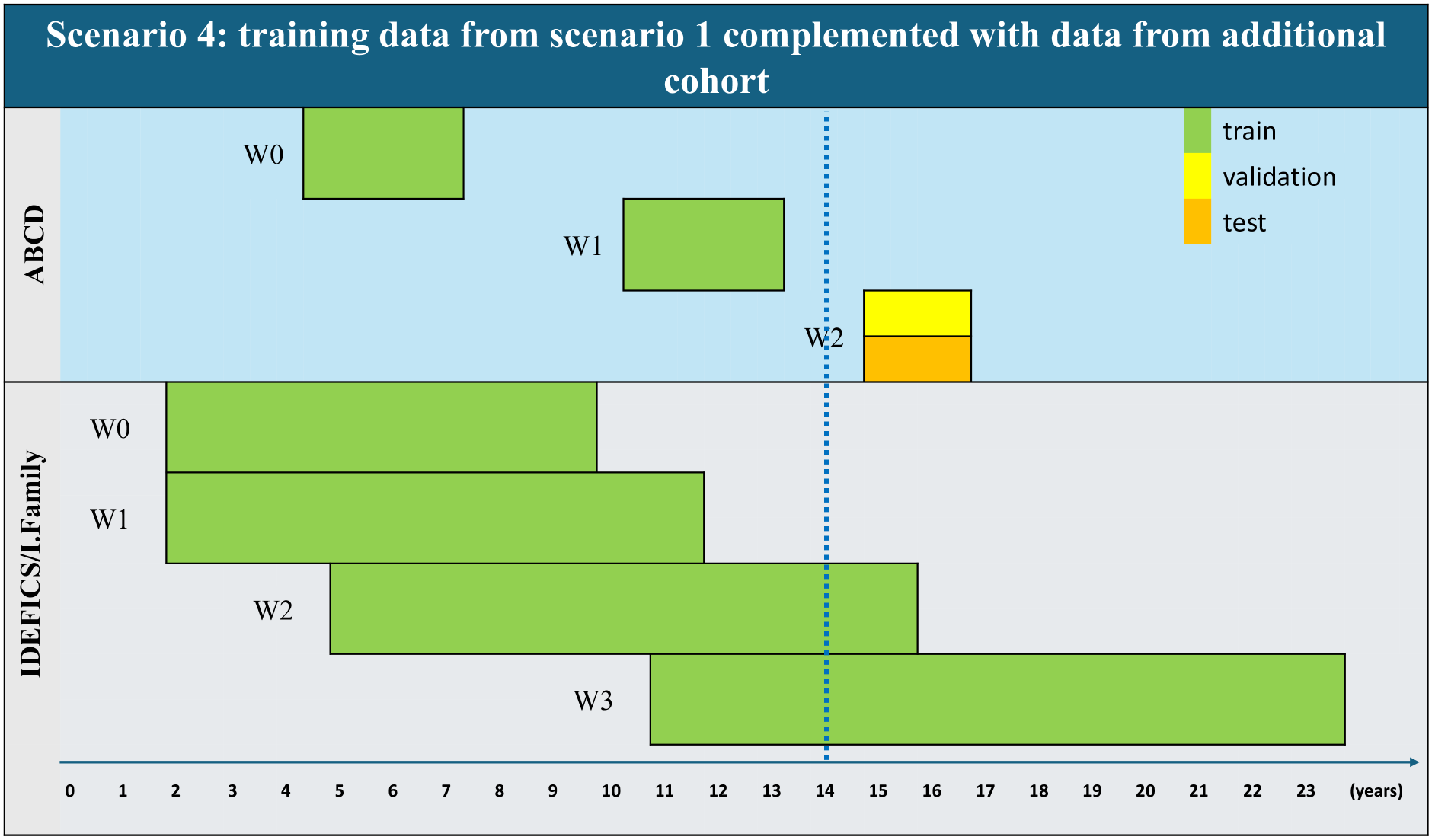}

  \caption{Multi-cohort scenarios}
  \label{f:design34}
\end{figure}

In our illustrative example, we predict BMI after age 14 for ABCD participants who have previous measurements before age 14. To assess the benefit of additional data from the second cohort, we design four scenarios of forecasting BMI z-score in participants seen in training using information from single-cohort data as shown in Figure~\ref{f:design12} and multi-cohort data with additional measurements at later ages as shown in Figure~\ref{f:design34}.

Scenario 1 mimics the common situation that longitudinal data are only available up to a certain age. ABCD observations are used only up to age 14 as training data. ABCD observations at ages older than 14 years (from 1730 subjects with measurements before age 14) are randomly split into validation (865 subjects) and test data (865 subjects). Scenario 2 represents the ideal situation that follow-up data taken at older ages from the same cohort are available in a subset. We randomly split the validation set from scenario 1 into five folds: each time 1/5 of the validation data is added to the training set while the other 4/5 is kept as validation data. In scenario 3, we augment the ABCD training set from scenario 1 with data from the IDEFICS/I.Family cohort, but to reduce heterogeneity we include only 3943 IDEFICS/I.Family participants from countries with similar demographic profile to the Netherlands (specifically, Germany and Belgium, considered ``central EU''). In scenario 4, data from all IDEFICS/I.Family participants are added to the ABCD training set from scenario 1, which introduces greater heterogeneity in the training data. 

Within each scenario, we consider the models listed in Table~\ref{tab:models}. For both LMM and MERF$_{multi-cohort}$, as the true DGP is unknown, we estimate different models with varying variance-covariance structures and select the one that best fits the validation data, using mean absolute error (MAE) as evaluation metric. In contrast, for standard MERF \citep{Hajjem2014}, we include the subject-level random effects and use the covariance structure assuming independent residuals within-subject and constant residual variance, because alternative structures are not supported by the available implementation. After selecting an appropriate structure, we compare the predictive performance of using the most recent outcome measurements before age 14 as predictions, LM, LMM, RF and standard MERF approach \citep{Hajjem2014} against the proposed MERF$_{multi-cohort}$. The comparison is based on the test data that are held-out and kept the same across scenarios.

\begin{table}[t]
\centering
\caption{Models considered for the evaluation and comparison}
\label{tab:models}

\renewcommand{\arraystretch}{1.15}
\setlength{\tabcolsep}{5pt}

\begin{tabularx}{\linewidth}{|
>{\raggedright\arraybackslash}p{2.8cm}|
>{\raggedright\arraybackslash}X|}
\hline

\textbf{Model} & \textbf{Description} \\
\hline

$\mathrm{Previous}$ &
For subject $i$, the prediction at time $t$, $\hat{y}_i(t)$,
is the latest available measurement $y_i(t-1)$. \\
\hline

$\mathrm{LM}$ &
Linear model with $\mathbf{X}\boldsymbol{\beta}$. \\
\hline

$\mathrm{LMM}$ &
Linear mixed model with $\mathbf{X}\boldsymbol{\beta}$ as fixed effects,
with a random intercept and random slope for age for each subject,
implemented in the \texttt{nlme} package \citep{nlme}. \\
\hline

$\mathrm{RF}$ &
Random forest implemented in the \texttt{ranger} package
\citep{Wright2017}, using the same set of features $\mathbf{X}$
with default hyperparameters. \\
\hline

$\mathrm{MERF}$ &
Mixed-effects random forest proposed by Hajjem et al.~\citep{Hajjem2014},
implemented in the \texttt{LongituRF} package \citep{LongituRF},
using the same set of features $\mathbf{X}$ and the same
hyperparameters as $\mathrm{RF}$. \\
\hline

$\mathrm{MERF}_{\text{multi-cohort}}$ &
Our proposed $\mathrm{MERF}_{\text{multi-cohort}}$,
using the \texttt{ranger} package \citep{Wright2017}
for estimating the fixed-effects function and the
\texttt{nlme} package \citep{nlme} for estimating the random effects,
with the same set of features $\mathbf{X}$ and the same
random-forest hyperparameters as $\mathrm{RF}$. \\
\hline

\end{tabularx}
\end{table}

\section{Results and Interpretation}
\label{s:results}

The results of the model selection procedure for the variance--covariance structures are provided in the online Supplementary Material B, including the candidate structures evaluated and their corresponding performance on the validation data. The selected variance--covariance structure (with smallest mean absolute error on validation data) for each model are reported in the online Supplementary Material B.

Table~\ref{tab:results} presents the performance of models without random effects (LM and RF), the standard mixed-effects model (LMM), previous MERF approach ~\citep{Hajjem2014} and MERF$_{multi-cohort}$, evaluated under four scenarios. 

\begin{table}[t]
\centering
\begin{threeparttable}

\caption{Model performance evaluated using mean absolute error on the
test set across four scenarios}
\label{tab:results}

\renewcommand{\arraystretch}{1.15}
\setlength{\tabcolsep}{5pt}

\begin{tabularx}{\textwidth}{
l
>{\centering\arraybackslash}X
>{\centering\arraybackslash}X
>{\centering\arraybackslash}X
>{\centering\arraybackslash}X
>{\centering\arraybackslash}X
>{\centering\arraybackslash}X
}
\toprule
\textbf{Scenario}
& \textbf{Previous}
& \textbf{LM}
& \textbf{LMM}
& \textbf{RF}
& \textbf{MERF}
& \textbf{MERF$_{\text{multi-cohort}}$} \\
\midrule

1 & 0.729 & 0.687 & \textbf{0.635} & 0.681 & 0.645 & \textbf{0.635} \\
2 & 0.729 & 0.652 & \textbf{0.585} & 0.663 & 0.626 & 0.605 \\
3 & 0.729 & 0.655 & \textbf{0.598} & 0.664 & 0.660 & 0.613 \\
4 & 0.729 & 0.742 & 0.704 & 0.663 & 0.675 & \textbf{0.619} \\

\bottomrule
\textit{}\end{tabularx}

\begin{tablenotes}
\footnotesize

\item Scenario 1: Training data include ABCD observations only up to
age 14 (ABCD$<14$).

\item Scenario 2: Training data from Scenario 1 are augmented with
additional observations from ABCD subjects older than 14 years who are
not included in the test set (ABCD).

\item Scenario 3: Training data from Scenario 1 are complemented with
data from a similar population
(ABCD$<14$ + IDEFICS/I.Family$_{\text{central-EU}}$).

\item Scenario 4: Training data from Scenario 1 are complemented with
data from an additional cohort that differs from the target cohort in
its demographic profile
(ABCD$<14$ + IDEFICS/I.Family).

\end{tablenotes}

\end{threeparttable}
\end{table}

In scenario 1 and scenario 2, LMM outperforms LM, and similarly both MERF and MERF$_{multi-cohort}$ outperform RF. This suggests that in single cohort scenarios the prediction of future observations for a given subject is improved by considering the random effect associated with that subject. 

Within each scenario under investigation, the LMM outperforms the LM, and similarly MERF$_{multi-cohort}$ achieves better predictive performance than RF and MERF, indicating that models accounting for a hierarchical structure in the data through more flexible covariance structures have better predictive performance than models that assume oversimplified covariance structures. 

Across all four scenarios, the baseline method shows identical performance, as the test set is fixed and the baseline prediction relies only on the most recent observed outcome measurement for each subject. Except for the baseline method, prediction performance across all other models improves when adding future data from the same cohort. In both, scenarios 3 and 4, all models show lower prediction performance compared with scenario 2. However, in scenario 3 (training data complemented with data from a similar population), the decrease in predictive performance is minor.

\subsection{Benefits of \texorpdfstring{MERF$_{multi-cohort}$}{MERF multi-cohort} in scenario with heterogeneous data}
The advantage of MERF$_{multi-cohort}$ over MERF becomes more evident in multi-cohort scenarios, demonstrating the benefit of allowing more flexible covariance structures for random effects and residuals in multi-cohort settings, which are not supported in MERF.  

MERF$_{multi-cohort}$ and LMM show comparable performance in single-cohort scenarios and in the multi-cohort scenario involving similar cohorts. While the predictive performance of MERF$_{multi-cohort}$ remains relatively stable across all scenarios, the advantage of MERF$_{multi-cohort}$ is most evident in scenario 4, where the additional cohort is more heterogeneous. This suggests that, while both MERF$_{multi-cohort}$ and LMM can capture complex correlation structures, MERF$_{multi-cohort}$ provides greater flexibility in modeling nonlinear relationships and complex interactions.
\subsection{Benefits of additional data from external cohorts}
While increasing the amount of training data is often expected to improve model performance, this improvement is not guaranteed when cohort heterogeneity is ignored. Our results suggest that the optimal scenario for predicting future BMI trajectories is to have observations at later ages from a subset of participants within the same cohort. However, in many epidemiological studies, measurements are often only available up to certain ages, which restricts the ability to predict trajectories at later ages. In our results, predictive performance of MERF$_{\text{multi-cohort}}$ decreases in scenarios 3 and 4 compared with scenario 2, but remains better than in scenario 1. This indicates that heterogeneity in multi-cohort data makes prediction more difficult, but, if the heterogeneity is adequately accounted for by the model, additional data from other cohorts may improve extrapolation to later ages by extending the observed period beyond the target cohort's age range. 

Despite these promising results, it should be noted that the observed differences in predictive performance across methods are relatively small, and the incremental gain over the baseline method from incorporating additional predictors and more complex modeling is limited. A possible
explanation is that the current evaluation is based on two European child cohorts with moderate heterogeneity, and the application considers tabular data with a relatively small number of predictors and an outcome with limited temporal variability. 

\section{Discussion and Conclusion}
\label{s:discuss}
In this study, we extend the previous MERF approach \citep{Hajjem2014} to multi-cohort settings and generalize MERF in three respects: (i) subject-level rather than observation-level out-of-bag (OOB) predictions are used as fixed effect estimates to better separate population-level and subject-specific variation; (ii) affine calibration of the fixed effect contribution provides more flexibility in correcting the systematic discrepancy in the fixed effect estimate to improve
covariance estimation; (iii) flexible variance--covariance structures for both random effects and residuals are estimated as parts of the LMM. 

The simulation results demonstrate that the proposed MERF$_{multi-cohort}$ retains predictive performance comparable to standard MERF under simple covariance structure, while providing more accurate estimates of the variance-covariance components. Improvements in predictive performance and covariance parameter estimation are observed in some settings with complex covariance structure. The benefits of allowing flexible covariance structures are more apparent under residual heteroscedasticity, which may arise in multi-cohort data from differences in measurement procedures and tools. 


We demonstrate MERF$_{multi-cohort}$ for predicting individual-level BMI z-score trajectories using pooled data from two children cohorts. MERF$_{multi-cohort}$ provides greater flexibility in modeling the covariance structure in the analysis of complex multi-cohort data, which is limited in previous MERF approaches. At the same time, MERF$_{multi-cohort}$ enables flexible modeling of fixed effects with non-linear patterns and complex interactions. This combined flexibility in modeling both the random effects and fixed effects components allows MERF$_{multi-cohort}$ to better leverage larger and more heterogeneous training data.

Our proposed approach separates the estimation of fixed and random effects. As a result, our proposed approach is not restricted to RF and in principle could be extended to other ML methods. This separation also allows the use of established explainability methods for the ML component, such as variable importance scores used for RF by Ngufor et al.~\citep{Ngufor}, which may help to retain interpretability despite the use of flexible ML models. Compared to standard MERF \citep{Hajjem2014}, our new approach could have more correct attribution of random effects and fixed effects, which may improve the interpretation. MERF$_{multi-cohort}$ is currently limited to continuous outcomes. Future work could extend the approach by integrating generalized mixed-effects models into RF, enabling the prediction of outcomes with non-normal distributions. 

Existing studies do not provide a rigorous convergence analysis of the iterative ME-ML estimation algorithm itself. Instead of analyzing the iterative procedure directly, Capitaine et al.~\citep{Capitaine2021} derive a result on the asymptotic prediction bias of the fitted MERF predictor by expressing the prediction bias as a linear transformation of the RF estimation bias and then invoking the RF consistency result of Scornet et al.~\citep{Scornet2015}. Similar to the approach of Capitaine et al.~\citep{Capitaine2021}, we derive in the Appendix that the asymptotic prediction bias of the fitted MERF$_{multi-cohort}$ predictor is inherited from the RF component when the variance components are considered as known. However, extending the asymptotic result of Capitaine et al.~\citep{Capitaine2021} to the proposed MERF$_{multi-cohort}$ requires further investigation because existing RF consistency results do not directly apply to the present setting, which involves heterogeneous covariance structures across cohorts and potentially remaining temporal dependence after including the random effects.

Additionally, model evaluation in longitudinal prediction settings that leverage previous measurements to predict future outcomes for the same subjects is challenging. Standard cross-validation strategies are not straightforward in such settings because random partitioning of the data at the observation level may break temporal dependencies and lead to information leakage between training and validation sets, whereas partitioning at the subject level removes prior measurements of the same subjects from the training data. Therefore, model performance is evaluated based on their MAE calculated on a heldout test set that preserves the temporal structure of the observations. Future work could further investigate how to quantify the uncertainty of the estimated prediction performance in such settings, for example by adapting advanced techniques such as the approach proposed by \citet{Rink2025} to these settings.

Overall, our study highlights the potential of ME-ML approaches for predicting individual-level health trajectories in multi-cohort studies by combining ML methods with an explicit modeling of complex hierarchical structures and cohort heterogeneity through flexible random effects and residual covariance structures. 
However, further evaluation is required to assess the performance of the proposed method in a wider range of data settings with varying numbers of cohorts, stronger heterogeneity across cohorts, outcomes showing greater temporal variability, stronger nonlinear relationships, and more complex data structures that cannot be easily modeled using LMM, where its advantages are expected to be more pronounced.


\section*{Acknowledgments}
This work was funded by the Deutsche Forschungsgemeinschaft (DFG, German Research Foundation) - 459360854 as part of the Research Unit "Lifespan AI: From Longitudinal Data to Lifespan Inference in Health'' (DFG FOR 5347), University of Bremen (http://lifespanai.de).

\section*{Conflicts of Interest}

The authors declare no conflicts of interest.

\section*{Supporting Information}

The R scripts for the simulation study can be downloaded from an Open Science OSF project accompanying this manuscript. See: https://osf.io/c39tv/overview. Data from the ABCD is not publicly available and can be obtained upon application to the respective cohort study. Access to and use of IDEFICS/I.Family cohort data and biosamples can be obtained via an electronic application portal at https://www.bips-institut.de/en/research/research-data-portal. R scripts for the real data example are provided in the OSF project. Online Supplementary Materials A and B, referenced in Section~\ref{s:MERF}, Section~\ref{s:setup} and Section~\ref{s:results}, are available with this paper and are also provided in the OSF project.\vspace*{-8pt}

\appendix

\section{Simulation\label{simulation}}

The simulated fasting-insulin trajectories are based on the the qualitative shape of the transient pubertal pattern: insulin sensitivity decreases after pubertal onset, reaches its lowest level around mid-puberty, and then recovers toward pre-pubertal levels by late puberty \citep{Moran1999}.

\subsection{Baseline and time-varying covariates\label{app1.1a}}
Let $i=1,\ldots,n_k$ denote subjects, $j=1,\ldots,n_i$
measurement occasions, $k$ cohort membership, and $t_{kij}$ the age of subject $i$ from cohort $k$ at measurement $j$. To allow distribution drift between cohorts, cohort-specific distributional differences in the covariates are introduced through
\[
\delta_k =
\begin{cases}
0, & k=C1,\\
\kappa, & k\neq C1.
\end{cases}
\]

Four baseline covariates were generated once per subject and kept constant across repeated measurements. Sex was generated according to
\[
S_i\mid k \sim
\operatorname{Bernoulli}\left(0.50+0.10\delta_k\right).
\]
The continuous baseline covariates $X_1$ and $X_2$ were jointly
generated as
\[
\begin{pmatrix}
X_{1i}\\
X_{2i}
\end{pmatrix}
\Bigg|k
\sim
\mathcal N_2
\left[
\begin{pmatrix}
1.5(1+\delta_k)\\
1.0(1+\delta_k)
\end{pmatrix},
\begin{pmatrix}
2+\delta_k & \rho_B\\
\rho_B & 1+2\delta_k
\end{pmatrix}
\right],
\]
where $\rho_B=0.3$. The third continuous baseline covariate was
generated from
\[
X_{3i}\mid k
\sim
\operatorname{Uniform}
\left(
-1+0.5\delta_k,\,
1+0.5\delta_k
\right).
\]

Two time-varying covariates were generated with continuous-time
within-subject correlation. For
\[
\mathbf X_{4i}
=
(X_{4i1},\ldots,X_{4in_i})^\top,
\]
we assumed
\[
\mathbf X_{4i}\mid k
\sim
\mathcal N_{n_i}
\left(
0.4\delta_k\mathbf 1_{n_i},
\boldsymbol{\Sigma}_{4,k}
\right),
\]
where
\[
[\boldsymbol{\Sigma}_{4,k}]_{jj'}
=
(1+\delta_k)^2
\left(\rho_4+0.1\delta_k\right)^{|t_{ij}-t_{ij'}|},
\]
with $\rho_4=0.2$.

For the second time-varying covariate,
\[
\mathbf X_{5i}
=
(X_{5i1},\ldots,X_{5in_i})^\top,
\]
we assumed
\[
\mathbf X_{5i}
\mid S_i,\mathbf {t_i},k
\sim
\mathcal N_{n_i}
\left(
\boldsymbol{\mu}_{5,i,k},
\boldsymbol{\Sigma}_{5,k}
\right),
\]
with
\[
\mu_{5,ij,k}
=
0.1S_i+0.1t_{ij}+0.4\delta_k,
\]
and
\[
[\boldsymbol{\Sigma}_{5,k}]_{jj'}
=
(1+\delta_k)^2
\left(\rho_5+0.1\delta_k\right)^{|t_{ij}-t_{ij'}|},
\]
where $\rho_5=0.2$.
\\

\subsection{True fixed effects function\label{app1.1b}}

The fixed-effect function included a smooth pubertal component to represent the transient increase in insulin resistance observed during puberty \citep{Moran1999}. The Gaussian form and its location and scale parameters were specified for simulation purposes to provide a smooth approximation to this transient pattern rather than to reproduce an empirically estimated age-specific trajectory.

\begin{equation}
P(a)
=
\exp
\left\{
-\frac{1}{2}
\left(
\frac{a-14}{2}
\right)^2
\right\},
\end{equation}

together with linear effects of baseline and time-varying covariates, 
a nonlinear adiposity effect, where steeper increase in fasting insulin is observed among subjects with overweight/obesity
\begin{equation}
\text{high\_adiposity}_i
=
\max\left(X_{4i}-0.5,\,0\right)^2
=
\begin{cases}
0, & X_{4i} \leq 0.5,\\[4pt]
(X_{4i}-0.5)^2, & X_{4i} > 0.5.
\end{cases}
\end{equation}

and interactions between baseline and time-varying covariates. Specifically,

\begin{equation}
\label{eq:true_fixed}
\begin{split}
f(\mathbf{X}_{igj},a_{igj})
={}&
\log(50)
+ 0.4P(a_{igj})
+ 0.1S_{ig}
+ 0.1X_{1,ig}
- 0.1X_{2,ig} \\
&+ 0.08X_{3,ig} \\
&+ 0.15X_{4,igj}
+ 0.15X_{5,igj}
+ 0.3\,\mathrm{high\_adiposity}_{i} \\
&+ 0.3X_{1,ig}X_{4,igj}
+ 0.3P(a_{igj})X_{4,igj}.
\end{split}
\end{equation}

Here, $S_{ig}$ denotes sex, $X_{1,ig}$ to $X_{3,ig}$ denote baseline covariates, 
and $X_{4,igj}$ and $X_{5,igj}$ denote time-varying covariates. 
\\

\section{Derivation of the asymptotic bias of the fitted \texorpdfstring{MERF$_{multi-cohort}$}{MERF multi-cohort} predictor}
\label{s:appendix}
Consider the MERF$_{multi-cohort}$ as a special case in the semi-parametric stochastic mixed effects model \eqref{eq:SMERF} in Section~\ref{s:MERF} proposed by \citet{Capitaine2021}, where the stochastic process $\boldsymbol{\omega}_i$ is not considered and flexible covariance structures for random effects and residuals are allowed:

\begin{equation}
\label{eq:model}
Y_i
=
f_i
+
Z_i b_i
+
\varepsilon_i,
\end{equation}

where

\[
b_i \sim \mathcal N(0,D_i),
\qquad
\varepsilon_i \sim \mathcal N(0,R_i).
\]
Then we derive the asymptotic bias of the fitted MERF$_{multi-cohort}$ predictor in the following way similar to the approach of \citet{Capitaine2021} assuming the variance-covariance
components are known.

The marginal covariance matrix of \(Y_i\) is

\begin{equation}
\label{eq:Vi}
V_i
=
Z_i D_i Z_i^\top
+
R_i.
\end{equation}

The fixed-effects component \(f_i=f(X_i)\) is estimated using random forests, yielding the estimator \(\widehat f_i\).

The fitted predictor is defined as

\begin{equation}
\label{eq:Yhat}
\widehat Y_i
=
\widehat f_i
+
Z_i \widehat b_i,
\end{equation}

where \(\widehat b_i\) is the BLUP estimator of the random effects expressed as:

\begin{equation}
\label{eq:bhat}
\widehat b_i
=
D_i Z_i^\top V_i^{-1}
\left(
Y_i-\widehat f_i
\right).
\end{equation}

Substituting \eqref{eq:bhat} into \eqref{eq:Yhat} gives

\begin{align}
\widehat Y_i
&=
\widehat f_i
+
Z_i D_i Z_i^\top V_i^{-1}
\left(
Y_i-\widehat f_i
\right).
\end{align}

Define

\begin{equation}
\label{eq:A}
A_i
=
Z_i D_i Z_i^\top
V_i^{-1}.
\end{equation}

Then

\begin{align}
\widehat Y_i
&=
\widehat f_i
+
A_i
\left(
Y_i-\widehat f_i
\right)
\\
&=
(I-A_i)\widehat f_i
+
A_i Y_i.
\end{align}

Taking expectations yields

\begin{equation}
\label{eq:EYhat1}
\mathbb E(\widehat Y_i)
=
(I-A_i)\mathbb E(\widehat f_i)
+
A_i \mathbb E(Y_i).
\end{equation}

From model \eqref{eq:model},

\begin{equation}
\label{eq:Truemean}
\mathbb E(Y_i)=f_i,   
\end{equation}

because

\[
\mathbb E(b_i)=0,
\qquad
\mathbb E(\varepsilon_i)=0.
\]

Moreover, define the random forest bias by

\begin{equation}
\label{eq:bias}
\operatorname{Bias}(\widehat f_i)
=
\mathbb E(\widehat f_i)-f_i.
\end{equation}

Hence,

\[
\mathbb E(\widehat f_i)
=
f_i+\operatorname{Bias}(\widehat f_i).
\]

Substituting into \eqref{eq:EYhat1}:

\begin{align}
\mathbb E(\widehat Y_i)
&=
(I-A_i)
\left(
f_i+\operatorname{Bias}(\widehat f_i)
\right)
+
A_i f_i
\\
&=
(I-A_i)f_i
+
(I-A_i)\operatorname{Bias}(\widehat f_i)
+
A_i f_i.
\end{align}

Since

\[
(I-A_i)f_i + A_i f_i = f_i,
\]

we obtain

\begin{equation}
\label{eq:mainbias}
\boxed{
\mathbb E(\widehat Y_i)-f_i
=
(I-A_i)
\operatorname{Bias}(\widehat f_i)
}.
\end{equation}

Using \eqref{eq:A}, this becomes

\begin{equation}
\boxed{
\mathbb E(\widehat Y_i)-f_i
=
\left[
I-
\left(
Z_i D_i Z_i^\top
\right)V_i^{-1}
\right]
\operatorname{Bias}(\widehat f_i)
}.
\end{equation}

If the RF estimator becomes asymptotically unbiased, then the mean fit of MERF prediction also tends to the true mean behavior $f_i$. Thus, under some regularity conditions on the random forest

\[
\operatorname{Bias}(\widehat f_i)\to0,
\]

then

\[
\mathbb E(\widehat Y_i)-f_i \to 0.
\]

This is the step where \citet{Capitaine2021} invoke the consistency result of \citet{Scornet2015}, which do not directly apply to multi-cohort setting due to the presence of heterogeneous covariance structures across cohorts, and the possibly remaining temporal dependence after including the random effects.

\clearpage
\bibliographystyle{unsrtnat}
\bibliography{MERFcohort}

\end{document}